\documentclass[manuscript=article]{achemso}

\setkeys{acs}{
abbreviations=true,
articletitle=true,
biblabel=plain,
chaptertitle=true,
doi=true,
email=true,
etalmode=truncate,
keywords=true,
maxauthors=100,
super=true
}

\setcitestyle{super,open={},close={}}

\usepackage[utf8]{inputenc}
\usepackage[T1]{fontenc}
\usepackage[english]{babel}
\usepackage[version=4]{mhchem}
\usepackage{siunitx}
\usepackage{graphicx}
\usepackage{longtable}
\usepackage{geometry}
\usepackage{caption}
\usepackage{float}
\usepackage{natbib}
\usepackage{setspace}
\usepackage{xkeyval}
\usepackage{array}
\usepackage{booktabs}
\usepackage{listings}
\usepackage{lmodern}
\usepackage{mathpazo}
\usepackage{microtype}
\usepackage{hyperref}
\hypersetup{breaklinks,colorlinks=true,linkcolor=black,filecolor=black,urlcolor=black,citecolor=black}

\usepackage{amsmath}
\usepackage{amssymb}
\usepackage{amsfonts}
\usepackage{latexsym}
\usepackage[compact]{titlesec}
\usepackage[usenames,dvipsnames]{xcolor}
\usepackage{mathptmx}
\usepackage{calc}
\usepackage{titletoc}
\usepackage{lipsum}
\usepackage{multirow}
\usepackage{amstext}
\usepackage{tabularx}
\usepackage{adjustbox}
\usepackage{epsfig}
\usepackage[section]{placeins}
\usepackage{natmove}
\usepackage{titlesec}
\usepackage{tablefootnote}
\usepackage{threeparttable}
\usepackage[d]{esvect}
\usepackage{mathtools}
\usepackage{physics}
\usepackage{textcomp}
\usepackage{subcaption}
\usepackage{soul}
\usepackage[none]{hyphenat}
\usepackage{microtype}

\vfuzz \hfuzz
\SectionsOn
\SectionNumbersOn
\AbstractOn

\hypersetup{colorlinks=true,citecolor=blue,urlcolor=blue}

\title{Flat bands in ultra-wide gap two-dimensional germanium dioxide}
\author{Rafael Franco Ribeiro Reis}
\author{Gabriel Elyas Gama Ara\'ujo}
\affiliation{Institute of Physics, Federal University of Goi\'as, Campus Samambaia, 74690-900, Goi\^ania, Goiás, Brazil}

\author{Danilo Kuritza}
\author{Jos\'e Eduardo Padilha}
\affiliation{Advanced Campus Jandaia do Sul, Federal University of Paran\'a, 86900-000, Jandaia do Sul, PR, Brazil}

\author{Alexandre Cavalheiro Dias}
\affiliation[UNB]{University of Bras{\'{i}}lia, Institute of Physics and International Center of Physics, Bras{\'{i}}lia $70919$-$970$, Federal District, Brazil}
\email{alexandre.dias@unb.br}

\author{Andr\'eia Luisa da Rosa}
\affiliation{Institute of Physics, Federal University of Goi\'as, Campus Samambaia, 74690-900, Goi\^ania, Goiás, Brazil}
\email{andreialuisa@ufg.br}
\author{Renato Borges Pontes}
\affiliation{Institute of Physics, Federal University of Goi\'as, Campus Samambaia, 74690-900, Goi\^ania, Goiás, Brazil}
\email{pontes@ufg.br}

\begin{document}

\begin{abstract}
We employ first principles density-functional theory (DFT) and the
Bethe-Salpeter equation (BSE) in the framework of tight-binding based
maximally localized Wannier functions (MLWF-TB) model to investigate
the electronic and optical properties of free-standing two-dimensional
(2D) germanium dioxide phases. All investigated 2D GeO$_2$ polymorphs
exhibit ultra-wide band gaps (3.6–5.3 eV) and strong excitonic
effects, with flat O-p-derived valence bands tunable under strain. These features allow the design of flat band
materials with ultra large electronic gaps in low-dimensional systems,
making these materials promising for devices operation at higher
voltages and temperatures than conventional semiconductor materials.
\end{abstract}

\maketitle

\section{Introduction}

Ultrawide bandgap (UWBG) semiconductors, with energy bandgaps larger than 4 eV, much wider than the conventional wide bandgap (WBG) of GaN (3.4\,eV) and SiC (3.2\,eV), represent an emerging new area of research covering a wide spectrum from materials, physics, devices, and applications. This new class of semiconductors has promising applications for future generations of high-power electronics, as well as deep-UV optoelectronics, quantum electronics, and extreme conditions applications.

Most of the research in ultra large band gap materials have been concentrated on conventional semiconductors such as  aluminum gallium nitride alloys\,\cite{D0TC04182C}, boron nitride\,\cite{aelm.202400549} and gallium oxide\,\cite{overview_ultra,overview_ultra2,aelm.202300844}.   The properties  of gallium oxide have been largely  investigated in the past years, but only recently this material has been considered for power electronics applications\,\cite{Mastro_2017}. The monoclinic $\beta$ phase of gallium oxide is reported to be thermally stable with  bandgap in the 4.6-4.9 eV range. Furthermore, its theoretically predicted critical field ranges more than twice the value of other wide bandgap semiconductors such as gallium nitride and silicon carbide\,\cite{Tadjer_2019}.  However, the absence of clear demonstrations of $p$-type doping in Ga$_2$O$_3$ makes it very difficult to achieve low turn-on voltages and ultra-high breakdown at the same time\,\cite{APR_2018}.

Aluminum nitride (AlN) and its alloys aluminum gallium nitride
(Al$_x$Ga$_{1-x}$N) possess a bandgap that can be tuned from 3.4 to
over 6\,eV by changing the Al composition $x$. This UWBG
semiconductors have potential applications in high-efficiency
optoelectronic devices, and ultra-high voltage power electronic
devices and deep UV
sensors\,\cite{aelm.201600501,10.1063/1.5138127}. In addition, AlN
crystals are also an ideal substrate material for epitaxial growth of
group-III nitrides, mitigating the shortcomings of large lattice
mismatch and large thermal mismatch with other
substrates\,\cite{Dalmau_2011}.  These properties make them
particularly suitable for fabricating surface acoustic wave resonators
and terahertz devices\,\cite{Bu2006}. Moreover, the high thermal
conductivity and high breakdown field of AlN materials are suitable
for high power field effect transistors\,\cite{pssa.201900694}.

Boron nitride (BN) is considered an ultra-wide bandgap (UWBG) semiconductor material with significant potential in various advanced electronic and optoelectronic applications\cite{EPJ2025}. Hexagonal boron nitride (h-BN), exhibits a large bandgap of approximately 6.0 eV and can withstand very high electric fields before breaking down, making them suitable for high-power devices. BN absorbs strongly UV light make it ideal for deep-ultraviolet photodetectors, light emitters, and lasers. BN high breakdown field and thermal conductivity make it a promising material for power switching devices. Furthermore, BN-based point defects can exhibit single-photon emission properties, making them relevant for quantum computing and cryptography\cite{10.1063/5.0021093}. However, producing high-quality p-type BN with high carrier mobility remains a challenge\,\cite{PhysRevB.96.100102}. Therefore, the search of suitable,  new materials could be an alternative to the already existing UWBG materials.

GeO$_2$ serves as a gate oxide in germanium-based Metal–Oxide–Semiconductor Field-Effect Transistors (MOSFETs), offering
superior carrier mobility compared to SiO$_2$ and is of critical importance in next-generation CMOS\cite{KAMATA200830}. Photodetectors and optoelectronic
devices based on thin GeO$_2$ films enhance stability and performance in Ge photodiodes
and waveguides\cite{CHOUDHURY2021111397}. However, compared to the development of conventional wide band
gap semiconductors, the understanding and development of ultra-wide
band gap materials is still in its infancy. Meanwhile, very thin
germanium oxide layers have appeared as interesting alternative owing
their ambipolar doping for application in high-power
devices\,\cite{10.1063/1.5088370}. Bulk germanium dioxide can
crystallize in several forms, with the two most common ones being the
hexagonal ($\alpha$-quartz-like, space group P3$_2$21) and tetragonal
(rutile-like, space group P4$_2$/mnm) phases. The hexagonal form has a
structure similar to $\alpha$-quartz, with germanium atoms in a
tetrahedral coordination\,\cite{GeO2_quartz}. The tetragonal phase has
a rutile-like structure, with germanium atoms in an octahedral
coordination\,\cite{geo2_rutile,acsanm.5c01137,10.1116/6.0002011}. GeO$_2$
can also exist in an amorphous form, which is similar to amorphous
SiO$_2$\,\cite{10.1021/acs.jpcc.6b07008}.  The $\alpha$-quartz phase
is the most stable phase of GeO$_2$ at room temperature and
atmospheric pressure and has an experimental band gap of 6.6 eV
\cite{acsanm.5c01137}.  Furthermore, germanium dioxide layers can also
be obtained in vitreous
form\cite{Lewandowski2019,acsomega3c05657,PhysRevB.95.155426,SHIN2023139851}. In
particular, several germanium dioxide structures were previously
predicted by a genetic algorithm \cite{PhysRevB.95.155426}.

In this work, we employ first-principles calculations within the
framework of Density Functional Theory (DFT) and a semi-empirical
approach grounded in the maximally localized Wannier function
tight-binding model (MLWF-TB) to solve the Bethe–Salpeter equation
(BSE) in order to investigate the electronic and optical properties of
2D germanium dioxide polymorphs. The germanium dioxide structures have
a very large electronic bandgap and remarbly well defined flat
bands. Our calculations reveal that excitonic effects are
pronounced, as demonstrated by the large exciton binding energy in
these materials. The combination of ultra-wide gaps, strong excitonic
interactions, and flat electronic bands places 2D-GeO$_2$ as
a promising platform for high-voltage, high-temperature, and
ultraviolet optoelectronic devices, as well as for exploring
correlation-driven phenomena in oxide-based 2D systems.

\section{Computational details} 

The calculations were performed employing density functional theory (DFT) within the generalized gradient approximation (GGA), explicitly employing the Perdew–Burke–Ernzerhof (PBE) exchange-correlation functional\,\cite{pbe} and the hybrid range-separated Heyd–Scuseria–Ernzerhof (HSE06) exchange-correlation functional\,\cite{hse}. The Kohn-Sham (KS) equations were solved using the projector augmented wave (PAW) method,\cite{paw} as implemented in the Vienna \textit{ab initio} Simulation Package (VASP)\,\cite{Kresse_13115_1993, Kresse_11169_1996}. Structure optimizations ensured the convergence of atomic forces to less than  0.01 eV/{\AA}.

To eliminate any spurious interactions between the monolayer and its periodic images in the $z$-direction, we incorporated a vacuum layer with a thickness of 15\,{\AA}. \textbf{k}-point meshes were automatically generated utilizing the Monkhorst-Pack method \cite{monkhorst1976special} with a sampling of $(6\times 6\times 1)$ {\bf k}-points. Relaxation and band structure calculations were performed using an energy cutoff of 450\,eV.

Ab-initio molecular dynamics (AIMD) simulations with the NVT ensemble at T= 300K (Nosé–Hoover thermostat) have been performed. The integration step of 1.0 fs during 10 ps has been considered. Additional AIMD calculations were performed for some selected geometries considering an integration step of 0.5 fs  and  16 ps. In this case, the initial time of 2.5 ps (5000 steps) was considered for the equilibration stage in all systems. Furthermore, phonon dispersion curves have been calculated using the finite displacement method\cite{phonopy1,phonopy2}. An energy cutoff of 520\,eV  was used and forces on atoms converged to less than 10$^{-6}$ eV/{\AA}. 

Linear optical response of the materials was computed to include excitonic effects using the Bethe-Salpeter equation (BSE) \cite{Salpeter_1232_1951} by employing the WanTiBEXOS package\cite{Dias_108636_2022}. A MLWF-TB Hamiltonian is created from first-principles DFT-HSE06 calculations employing  the Wannier90 package\cite{wannier90}. We have considered the  $s$ and $p$ orbital projections for \ce{Ge} and \ce{O} atoms. For the solution of the Bethe-Salpeter equation (BSE) \cite{Salpeter_1232_1951}, we employed the 2D Coulomb Truncated Potential (V2DT) \cite{Rozzi_205119_2006}. Our calculations were performed with a \SI{120}{\per\angstrom} \textbf{k}-points density for the periodic direction of the monolayers. We have also used a smearing of \SI{0.05}{\electronvolt} to determine the real and imaginary parts of the dielectric function. 

The production of images have been provided by both VESTA \cite{VESTA} and gnuplot.\cite{Gnuplot_4.4} Phonon calculations were performed using the density-functional perturbation theory (DFPT) method \cite{PhysRevB.43.7231,PhysRevB.55.10355} with additional support of Phonopy. \cite{phonopy1,phonopy2}. Additional information on computational parameters is available in the Support Information.

\section{Results and discussion}

Geometry is one of the key parameters for the electronic band structure of materials, which can lead to  exotic band structureS\,\cite{PhysRevLett.134.076402}. 
The optimized geometries are shown in Fig. \ref{fig:structure} for  (a) P-3m1, (b) P6/mmm, {c) Pbcm, (d) Pmma, (e) P4m2(1), (f) P4m2(2), (g) Pmmm(1) and (h) Pmmm(2) phases.  In particular, we investigate structures with rectangular, squared and hexagonal symmetries. Some of these structures were predicted by a genetic algorithm in Ref. \cite{PhysRevB.95.155426}.

\begin{figure*}[h]
\includegraphics[width=0.9\linewidth]{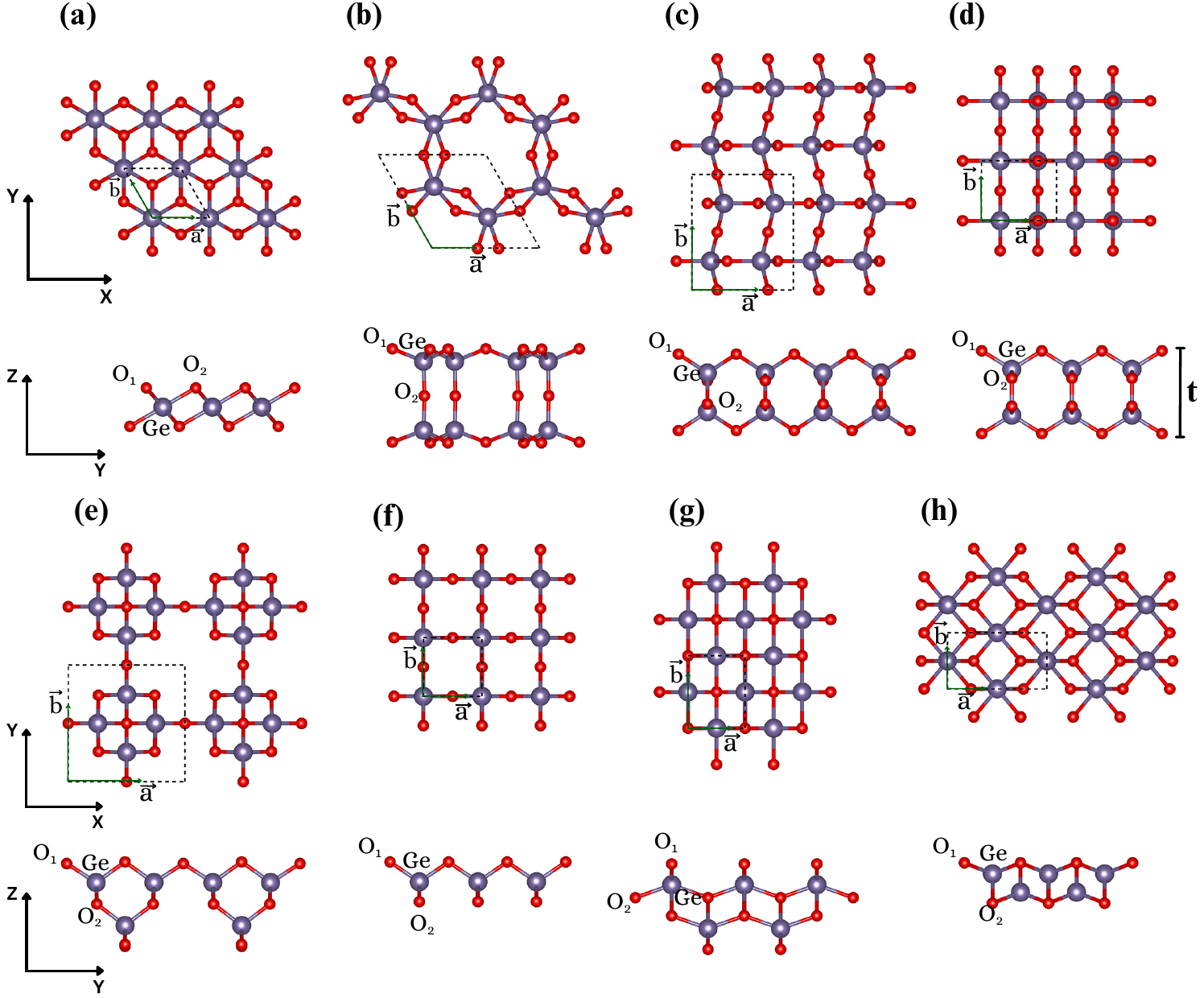}
\caption{Ball-and-stick representations of the optimized geometries of germanium dioxide layers within DFT-GGA for a) P-3m1, b) P6/mmm, c) Pbcm, d) Pmma, e) P4m2(1), f) P4m2(2), g) Pmmm(1) and h) Pmmm(2) phases. Red (gray) spheres represent oxygen (germanium) atoms. The unit cells are shown in dashed lines.}
\label{fig:structure}
\end{figure*}

\begin{table}[H]
\begin{tabular}{lcccccccc}
\hline
Space group & f.u. & $a$ (\AA)& $b$ (\AA) & $t$ (\AA)  & d$_{\rm Ge-O_1}$ ({\AA}) &  d$_{\rm Ge-O_2}$  ({\AA}) &  $\Delta H_f$(eV)\\\hline
P-3m1  & ${\rm GeO_2}$ & 2.92 & 2.92 & 1.96 & 1.95 & 1.95 & -3.08\\
P6/mmm & ${\rm Ge_4O_8}$ & 5.50 & 5.50 &  4.82 & 1.78 & 1.77 & -2.96  \\
Pbcm   & ${\rm Ge_4O_8}$ & 5.18 & 5.90 &  3.91 & 1.79  & 1.79  &  -3.26\\  
Pmma   & ${\rm Ge_2O_4}$ & 3.83 & 3.07 &  4.23 & 1.80 & 1.97  & -2.18\\
P4m2(1) & ${\rm Ge_4O_8}$ & 5.96 & 5.96 & 4.31 & 1.78 & 1.80  & -3.26\\
P4m2(2) & ${\rm GeO_2}$ & 2.99 & 2.99 &  2.03 & 1.80 & 1.80  & -3.14\\
Pmmm(1) & ${\rm Ge_2O_4}$ & 2.91 & 3.70 &  4.38 & 1.80 & 1.96 &  -2.96\\
Pmmm(2) & ${\rm Ge_2O_4}$ & 5.08 & 2.86 & 2.06 & 1.95 & 2.00 &  -2.30\\
\hline
\end{tabular}
\caption{Space group, formula unit, lattice parameters $a$ and $b$, thickness $t$, bond lengths ${\rm Ge-O_1}$, ${\rm Ge-O_2}$ and formation enthalpy $\Delta H_f$ of germanium dioxide layers.}
\label{tab:parameters}
\end{table}

In Table\,\ref{tab:parameters} the optimized inequivalent bond lenghts Ge-O$_1$  and Ge-O$_2$ vary between 1.78-1.95 {\AA} and 1.77-2.00\,{\AA}, respectively. For comparison, in the bulk rutile structure, each germanium atom is surrounded by five oxygen atoms with Ge-O bond lengths of 1.77-1.97{\AA}\cite{s41467-023-42890-3,s43246-022-00290-y}. The layer thickness $t$ varies from 1.96 {\AA} for P-3m1 and 4.82 {\AA} for P6/mmm. 
In particular, our results for in-plane lattice constants of the primitive unit cell, $a = b$; the atomic distance between neighboring Ge and O atoms, d$_{\rm Ge-O_{1,2}}$ and the thickness, t, for P-3m1, of: 2.92 \AA, 1.95 \AA\ and 1.96 \AA, are in close agreement with the results obtained by Sozen et. al \cite{D1CP02299G} of 2.91 \AA, 1.95 \AA\ and 1.97 \AA for a similar crystal lattice (1T-\ce{GeO2}). In addition, the calculated lattice parameter of 5.50 \AA\ for the P6/mmm structure  is in good agreement with the value of 5.49 \AA\ reported by Fuhrich et. al \cite {Fuhrich} for free-standing hexagonal \ce{GeO2} by using a similar scheme. They also showed, in accordance with our Fig 1 (b), that the P6/mmm most energetically stable structure is not formed by stacking oxygen atoms on top of each other. Instead, it is characterized by a network of \ce{GeO4} tetrahedra linked by bridging oxygen atoms.

The formation enthalpies at T = 0 K have been calculated as $\Delta H_f =  E_{total} - (N_{Ge}E_{Ge} + N_{O}E_{O})$, where ${\rm E_{total}}$, ${\rm E_{Ge}}$ and ${\rm E_O}$ are the total energies of the phase, germanium bulk in the diamond structure and oxygen molecule. ${\rm N_{Ge}}$ and ${\rm N_O}$ refer to the number of \ce{Ge} and \ce{O} atoms per unit cell.  The experimental enthalpy of formation of tetragonal bulk GeO$_2$ $\Delta H_f$(eV) is -5.80 eV\,\cite{JChemThermodynamics}.  We find that for germanium dioxide layers the enthalpy of formation  lies between -2.18 and -3.26 eV. With exception of Pmma, the formation enthaply of germanium dioxide layers lies around 3 eV.  As a matter of comparison Ref\,\cite{10.1116/6.0002011} reports -5.49\,eV for rutile GeO$_2$.

The phonon dispersion calculated at GGA-PBE is shown in Fig:\ref{fig:phonons}.  The dynamic stability of the systems was investigated via phonon dispersion by combining VASP and the Phonopy package\,\cite{phonopy1,phonopy2} based on the DFPT method.  The structures show small imaginary frequencies which indicates their dynamic stability at low temperatures.  Phonon dispersion calculations do not show a linear acoustic branch near the $\Gamma$ point for P6/mmm, P4m2(1), Pmmm(1), and Pmmm(2), which indicates bond softening and, therefore, structural instabilities for those phases in a free-standing form. However, a P6/mmm crystal lattice has already been experimentally obtained on several substrates \cite{Fuhrich,Lewandowski,PhysRevB.100.241403}, where epitaxial strain and electronic interactions with the substrate can stabilize phases that are otherwise unstable in their free-standing form.
In order to determine the structural stability of the  P6/mmm phase on Pt(111)(2x2) we have constructed an interface, where the surface lattice parameters of platinum is 11{\AA} in a  slab model containing 5 Pt layers. This leads to a lattice parameters mismatch of 3.2\%. We see from Fig. S12 that the P6/mmm structure does not undergo amorphization or strong distortions. 

\begin{figure}[H]
\includegraphics[width=0.6\linewidth,clip = true]{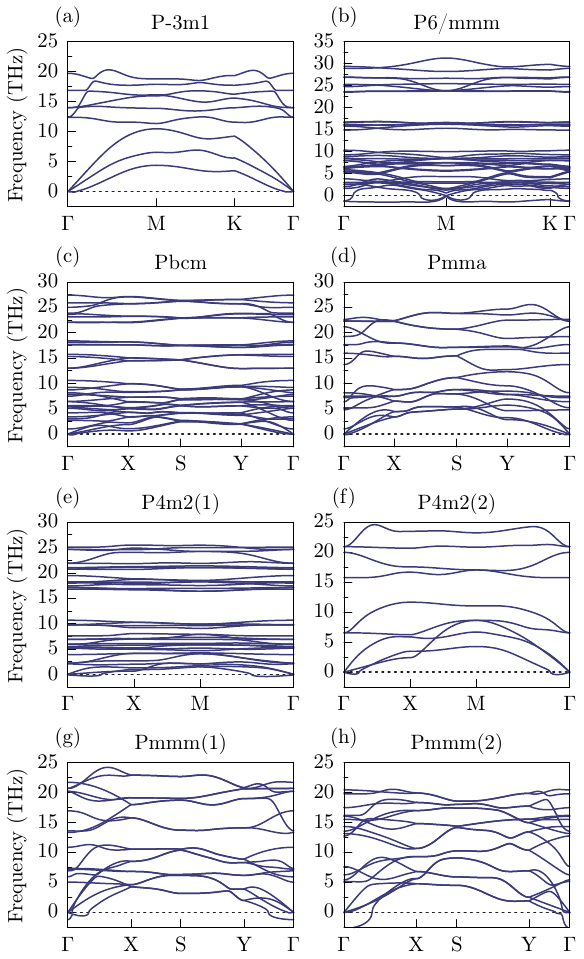}
\caption{Phonon dispersion curves for germanium dioxide layers: a) P-3m1, b) P6/mmm, c) Pbcm, d) Pmma, e) P4m2(1), f) P4m2(2), g) Pmmm(1) and h) Pmmm(2).}
\label{fig:phonons}
\end{figure} 

In addition to phonon spectra calculations, we performed AIMD simulations at T = 300 K for the 2D germanium dioxide structures considered in this work, with the results shown in Fig.\,\ref{fig:aimd}. A snapshot at the final step at 10\,ps simulation time with 0.5\,fs simulation step is shown. In addition to this simulation we have performed MD calculations for selected structures with additional simulation time of 16\,ps as shown in Fig.S2. With exception of Pmmm(2) all structures are thermally stable at 300K. We notice that the Pmmm(2) layer shows a strong bond distortion. However, as discussed before, it is possible that free standing 2D-germanium dioxide can be stabilized by a proper substrate, such as on ruthenium support as demonstrated in Ref.\,\cite{Lewandowski}.

\begin{figure}[H]
\includegraphics[width=0.8\linewidth]{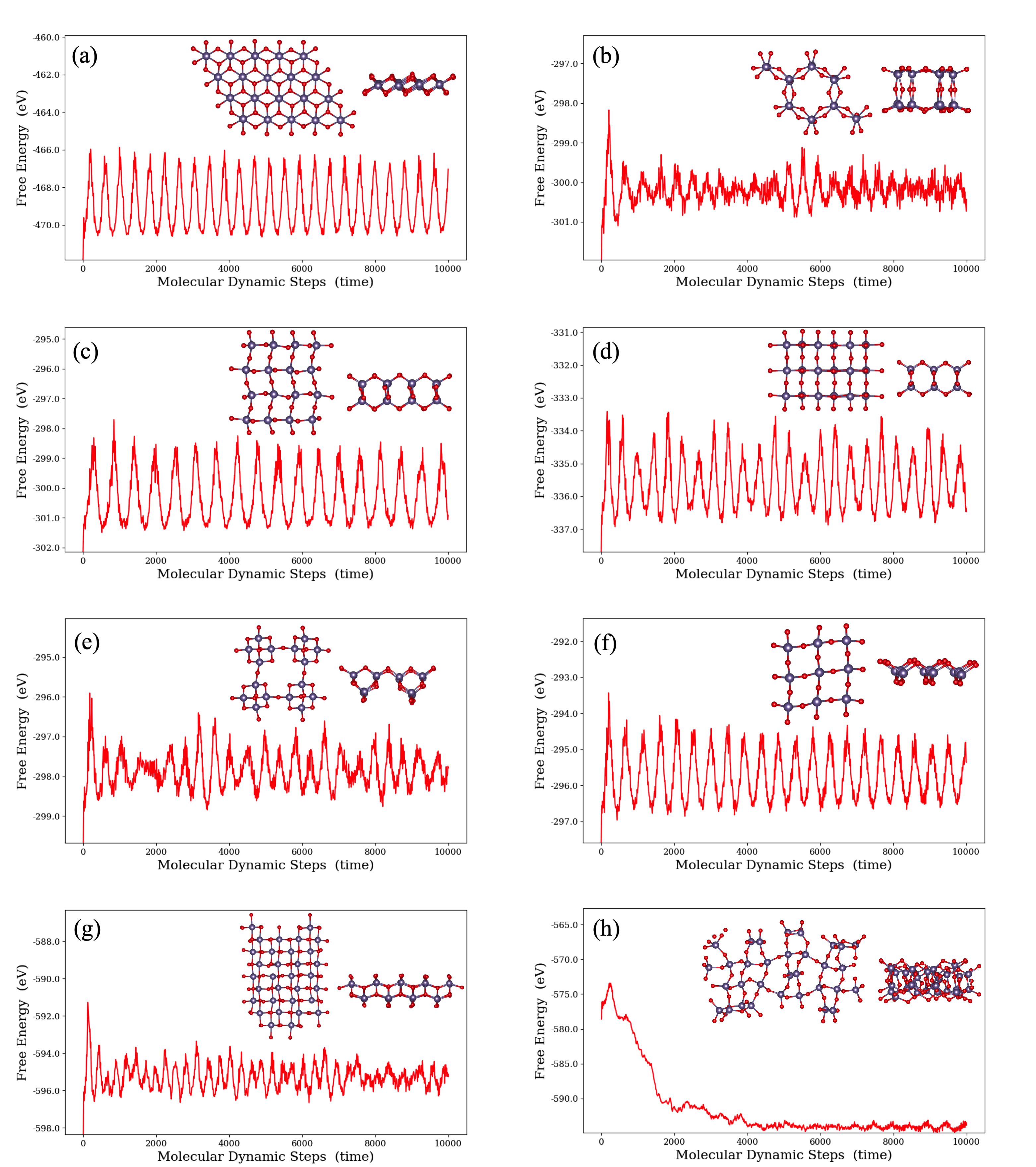}
\caption{AIMD calculations at 300K for a) P-3m1, b) P6/mmm, c) Pbcm, d) Pmma, e) P4m2(1), f) P4m2(2), g) Pmmm(1) and h\
) Pmmm(2). The snapshots  at 10\,ps simulation time are shown.}
\label{fig:aimd}
\end{figure}

\begin{figure}[H]
\centering
\includegraphics[width=0.9\linewidth]{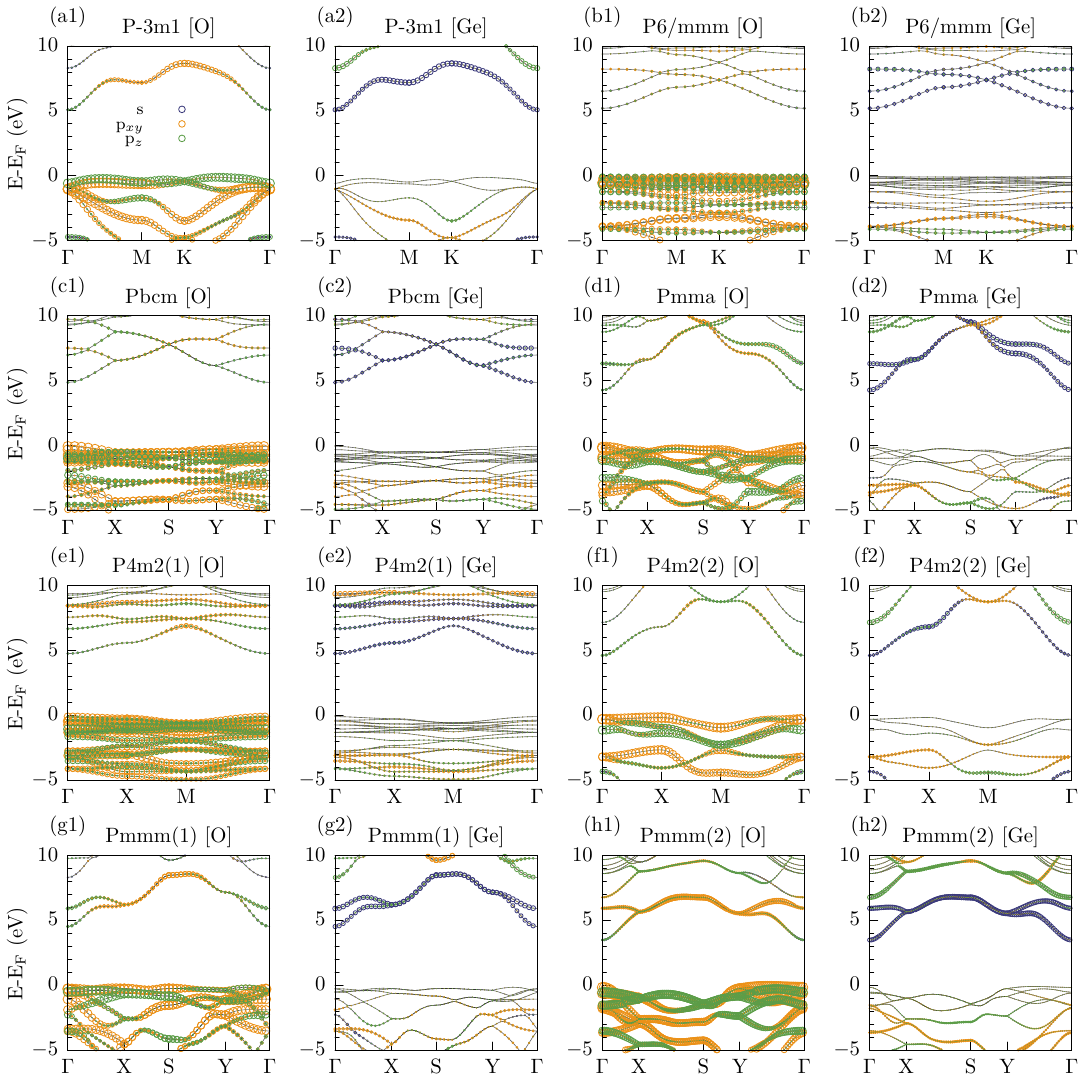}
\caption{Orbital projected band structure of germanium dioxide layers. a) P-3m1, b) P6/mmm, c) Pbcm, d) Pmma, e) P4m2(1), f) P4m2(2), g) Pmmm(1) and h) Pmmm(2). The Fermi levels are set as zero.}
\label{fig:pbands}
\end{figure}

Our first-principles calculations, employing the DFT-HSE06 hybrid
functional, reveal distinct electronic features for different
polymorphic phases of 2D germanium dioxide (\ce{GeO2}), as detailed in
the orbital-resolved band structures in Fig. \ref{fig:pbands}. A
universally observed feature is the presence of remarkably flat
valence bands across the entire Brillouin Zone, indicating a high
density of states and charge carrier localization. The orbital
analysis demonstrates that the valence band maximum (VBM) is
consistently dominated by O-$p$ orbitals, while the conduction band
minimum (CBM) predominantly originates from germanium s orbitals. The
P6/mmm, Pbcm,Pmma and P4m2(1) phases exhibit a direct band gap at the
$\Gamma$ point, with the VBM derived primarily from O p$_{\rm z}$
orbitals. In contrast, the P-3m1 P4m2(2),Pmmm(1) and Pmmm(2) phases
present an indirect band gap, with the VBM being a hybridization of O
p$_{\rm xy}$ and O p$_{\rm z}$ orbitals. Furthermore, van Hove
singularities emerge in the conduction band at the M point for
hexagonal symmetries (P-3m1, P6/mmm) and at the X point for the
tetragonal and orthorhombic symmetries (P4m2(1), Pbcm, Pmmm(1)). The
existence of multiple Dirac crossings well below the VBM is also
identified, suggesting the potential for emergent massless Dirac
fermions. These structural and electronic distinctions highlight the
broad tunability of the electronic properties of \ce{GeO2} monolayers
for applications in advanced electronic devices where electronic
correlation, magnetism, or excitonic condensation can be further explored\,\cite{CHOUDHURY2021111397}.

To have further insight into the the orbital-projected band structures presented in Fig.~\ref{fig:pbands}, the three-dimensional (3D) electronic band surfaces depicted in Fig.~\ref{fig:3dbands} offer a spatially resolved view of the energy dispersion near the band edges for eight distinct two-dimensional \ce{GeO2} polymorphs: P-3m1, P6/mmm, Pbcm, Pmma, P4m2(1), P4m2(2), Pmmm(1) and Pmmm(2). In each panel, the energy dispersion of the highest valence band (depicted in red) and the lowest conduction band (depicted in yellow) is plotted as a function of the in-plane wave vectors ($k_x$, $k_y$), centered around the $\Gamma$ point or other relevant high-symmetry points in the Brillouin zone. This representation complements the 1D band structure projections by offering a more comprehensive picture of band curvature, anisotropy, and gap alignment across the 2D reciprocal space. Notably, structures such as P-3m1 and P4m2(1) exhibit highly dispersive conduction bands with pronounced curvature, suggesting low electron effective masses and enhanced carrier mobility. Conversely, flatter bands in structures like P6/mmm and Pmmm(2) imply heavier carriers and more localized states.  These 3D electronic band plots, thus provide addtional insight into the electronic anisotropy and transport potential of each polymorph, strengthening the understanding of their suitability for applications in 2D nanoelectronic and optoelectronic devices. As a matter of completeness we show in the Fig. S9 the projected band structure of P-3m1 including Ge-$d$ orbitals in the band structure. One sees negligible difference, since Ge-$d$ orbitals lie very high in energy within the band structure.

\begin{figure}[H]
\centering
\begin{subfigure}[b]{0.24\linewidth}
\subcaption{}
\includegraphics[width=\linewidth,clip=true]{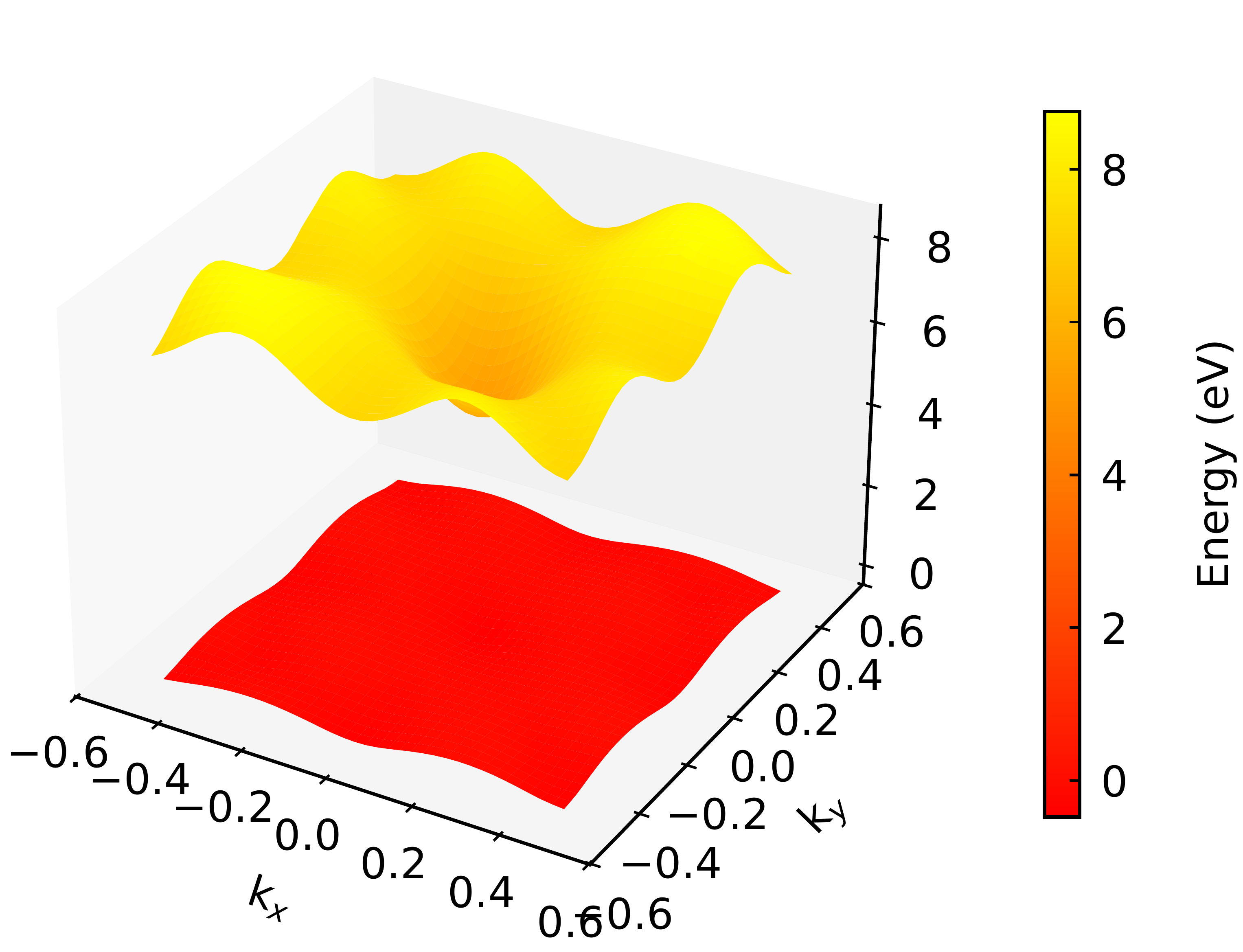}
\label{}
\end{subfigure}  
\begin{subfigure}[b]{0.24\linewidth}
\subcaption{}
\includegraphics[width=\linewidth,clip=true]{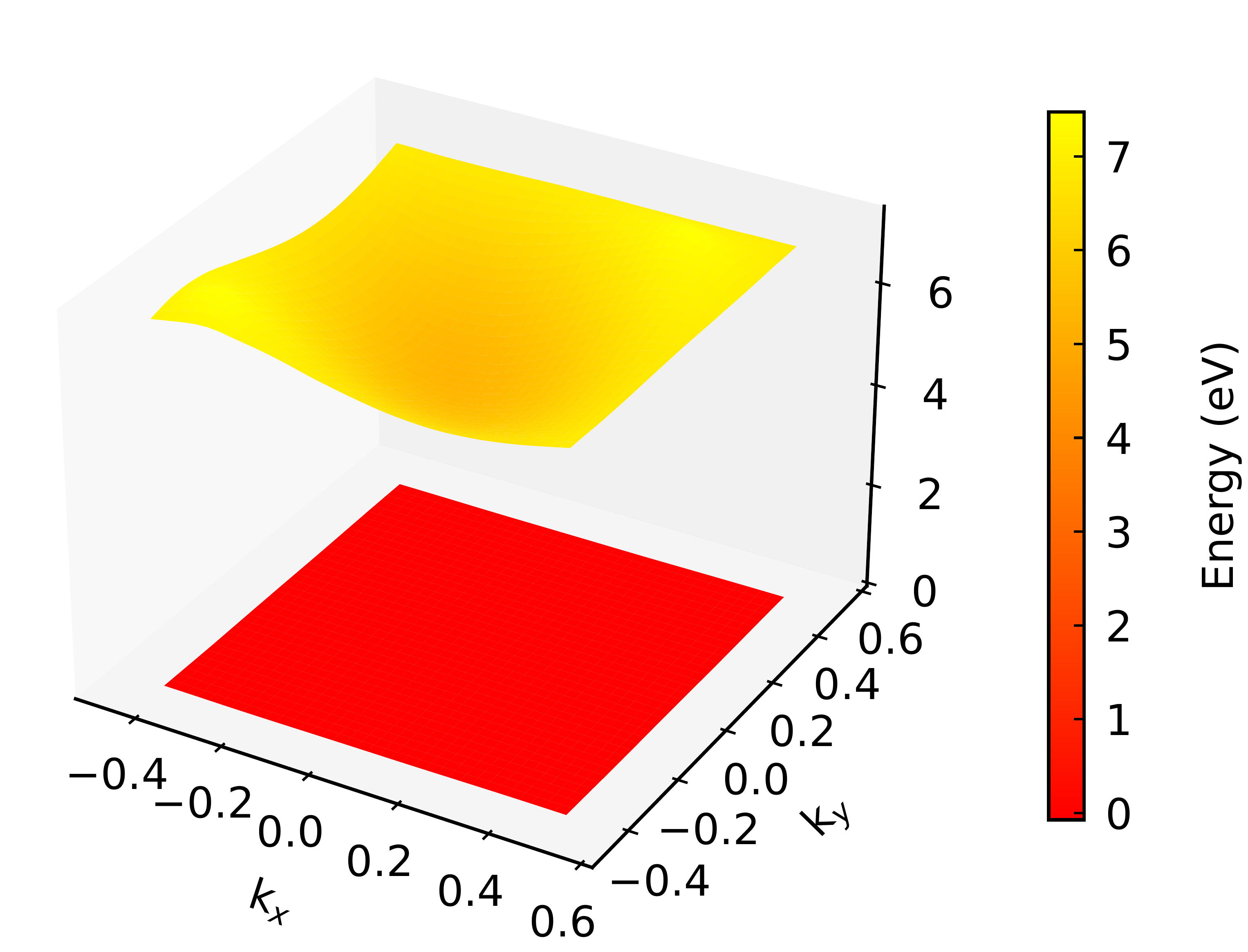}
\label{}
\end{subfigure}
\begin{subfigure}[b]{0.24\linewidth}
\subcaption{}
\includegraphics[width=\linewidth,clip=true]{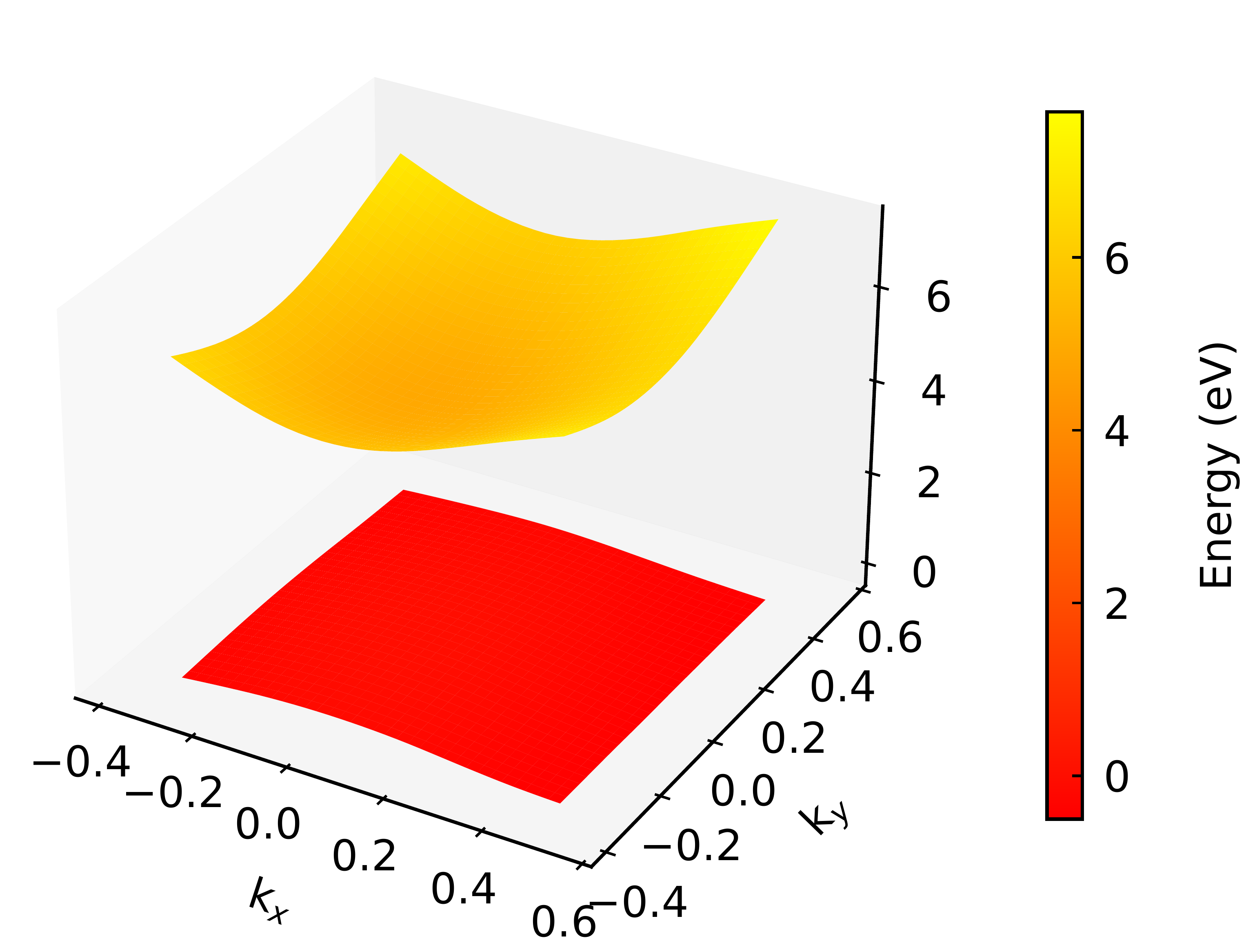}
\label{}
\end{subfigure}
\begin{subfigure}[b]{0.24\linewidth}
\subcaption{}
\includegraphics[width=\linewidth,clip=true]{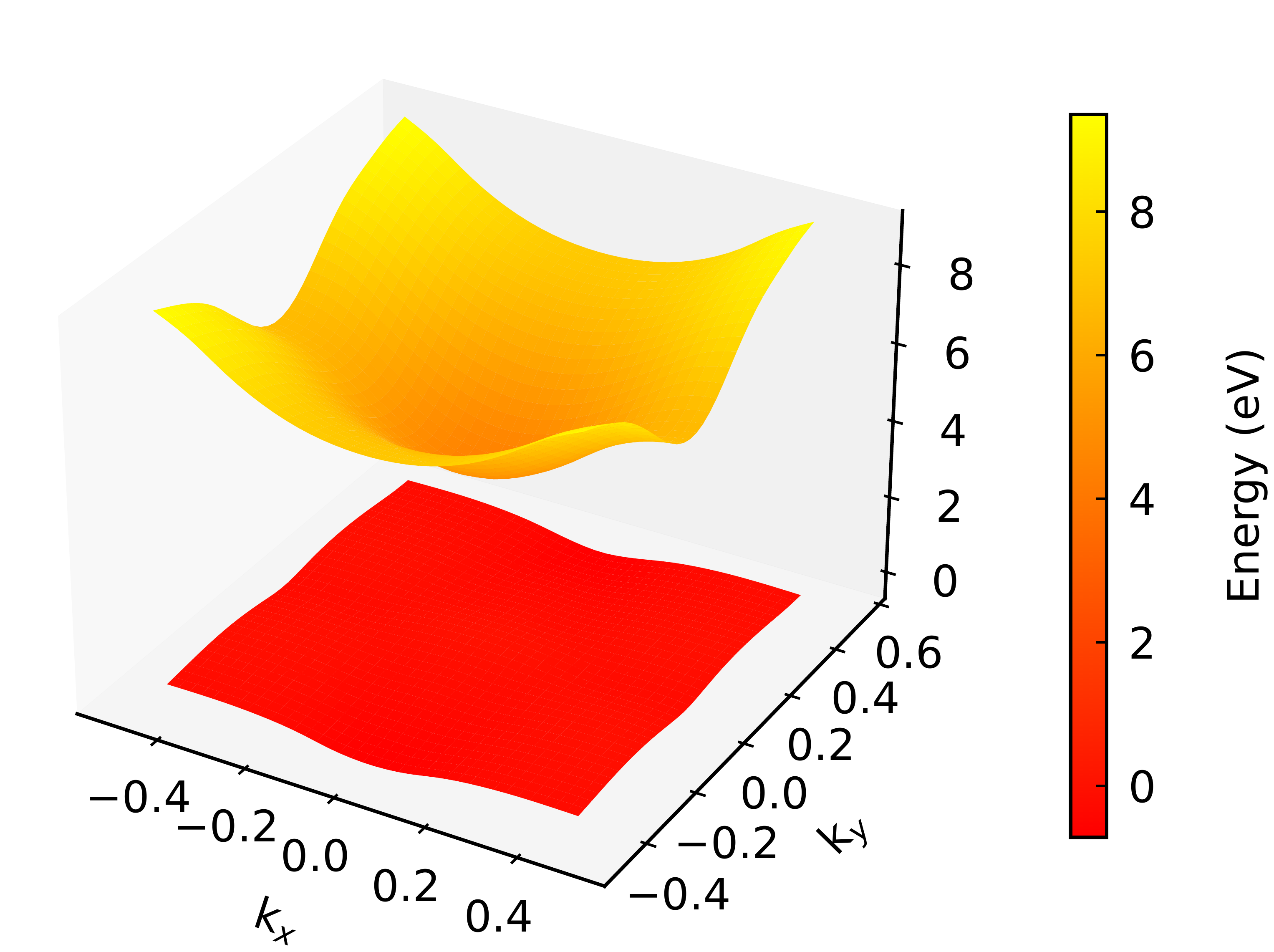}
\label{}
\end{subfigure}
\begin{subfigure}[b]{0.24\linewidth}
\subcaption{}
\includegraphics[width=\linewidth,clip=true]{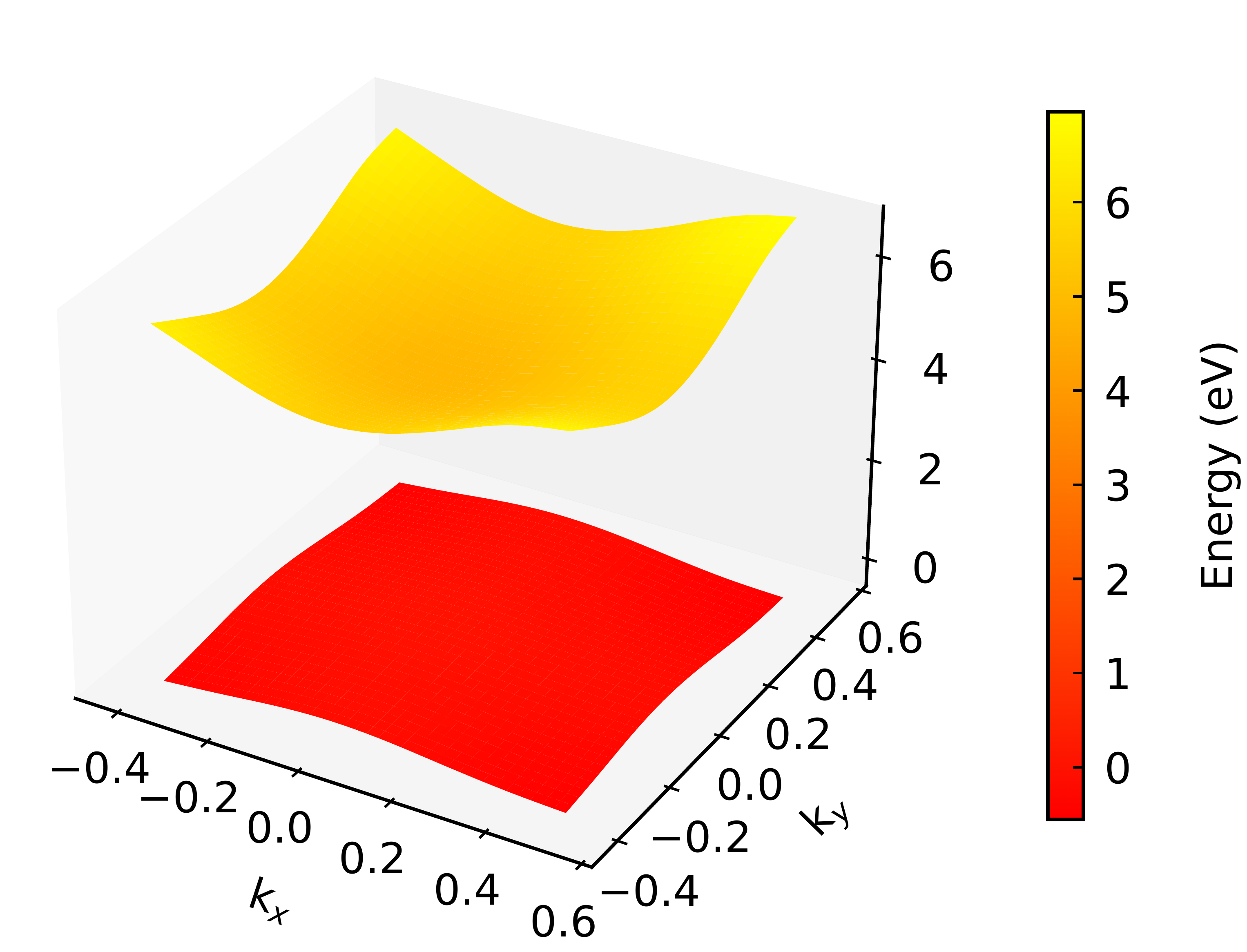}
\label{}
\end{subfigure}
\begin{subfigure}[b]{0.24\linewidth}
\subcaption{}
\includegraphics[width=\linewidth,clip=true]{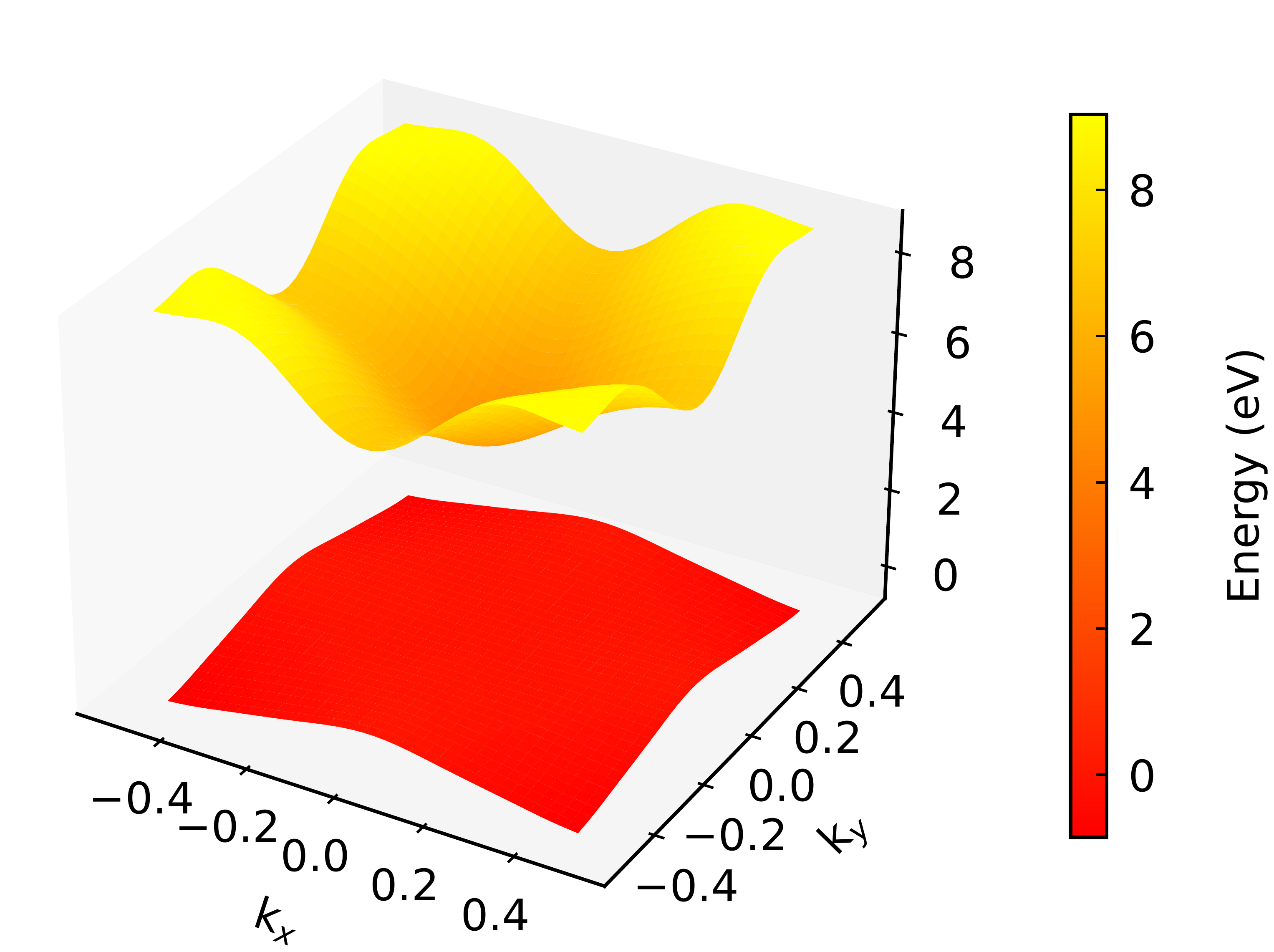}
\label{}
\end{subfigure}
\begin{subfigure}[b]{0.24\linewidth}
\subcaption{}
\includegraphics[width=\linewidth,clip=true]{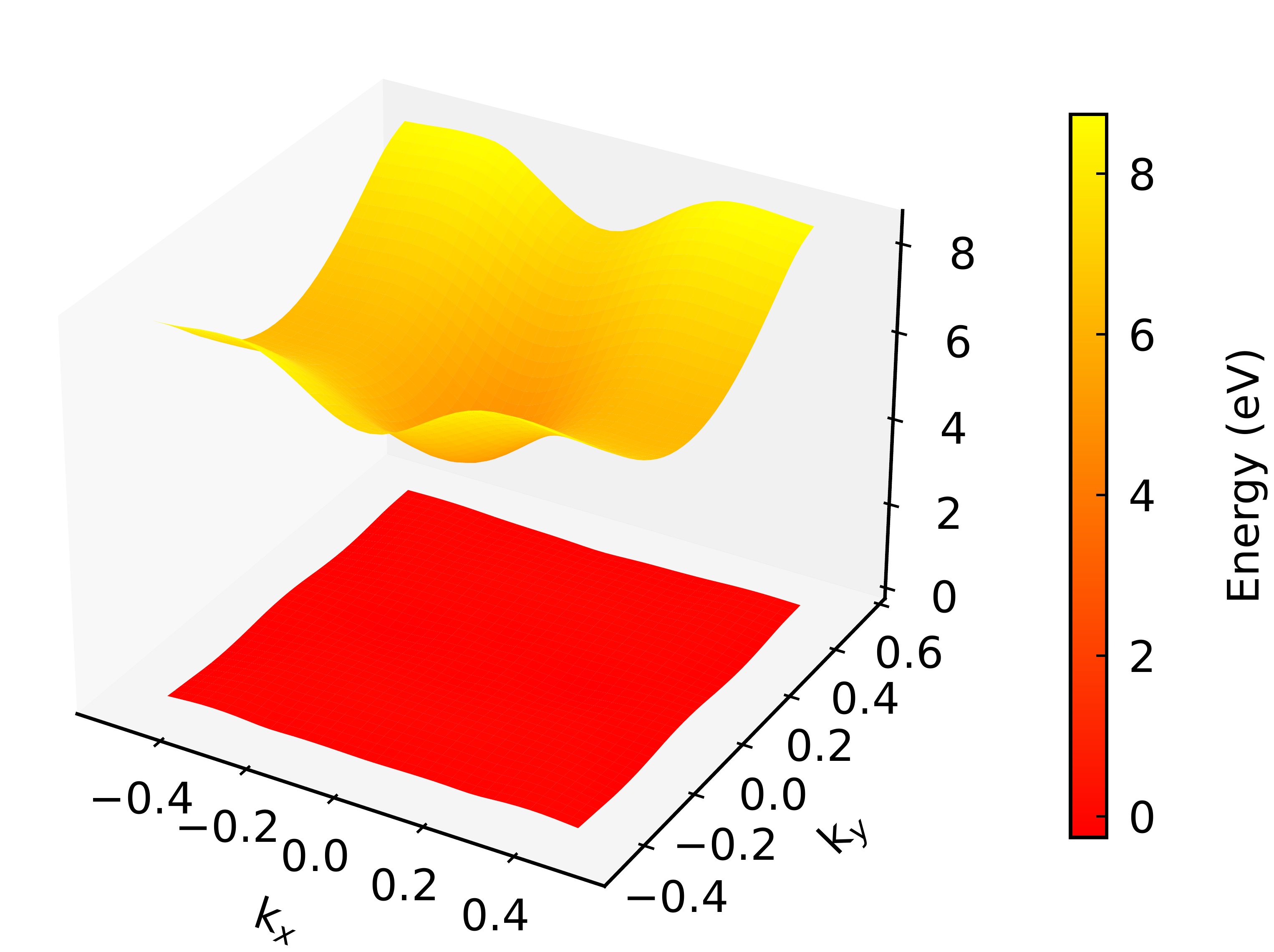}
\label{}
\end{subfigure}
\begin{subfigure}[b]{0.24\linewidth}
\subcaption{}
\includegraphics[width=\linewidth,clip=true]{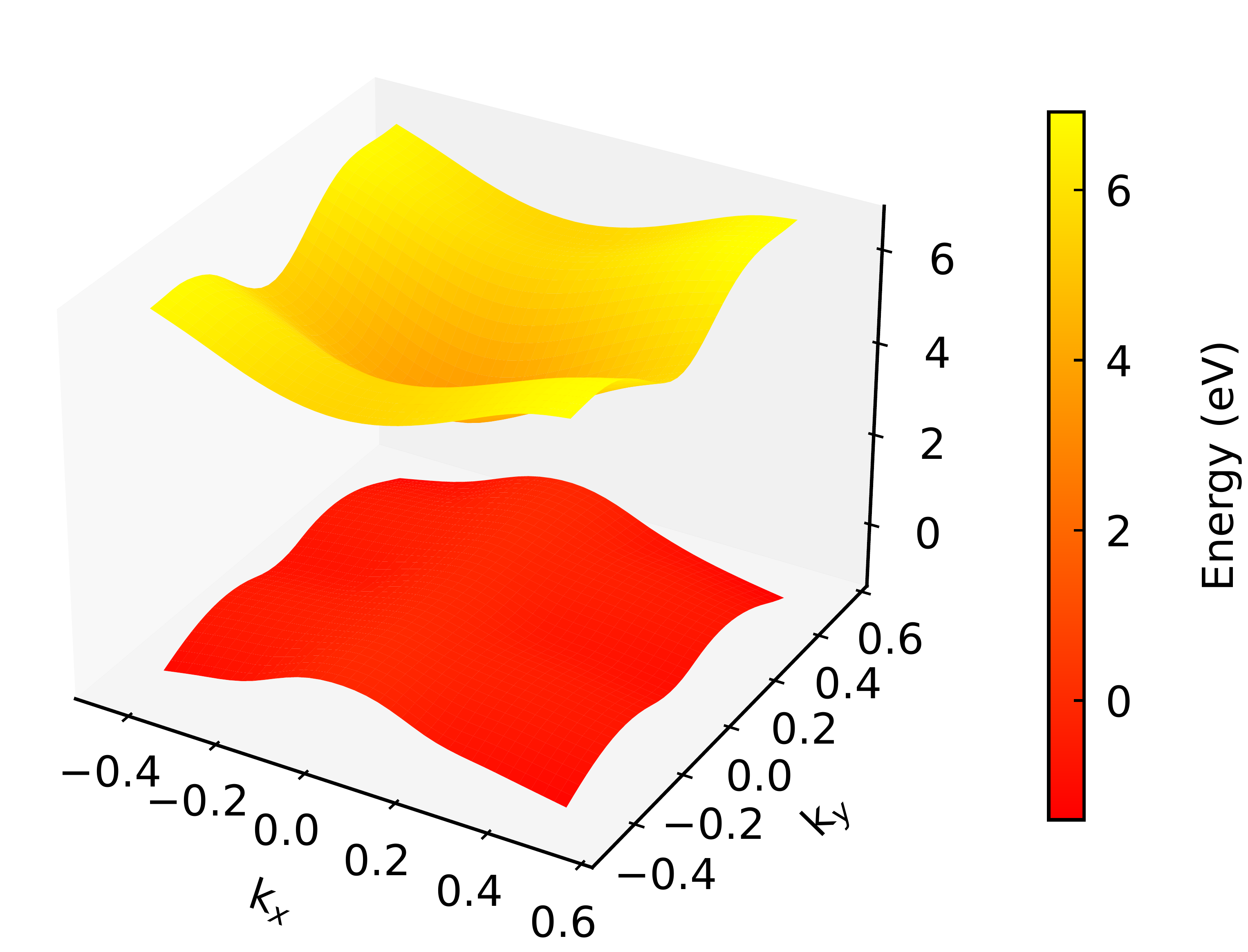}
\label{}
\end{subfigure}
\caption{Three-dimensional electronic band structures for 2D germanium oxide phases: a) P-3m1, b) P6/mmm, c) Pbcm, d) Pmma, e) P4m2(1), f) P4m2(2), g) Pmmm(1) and h) Pmmm(2).  The highest valence band (red) and the lowest conduction band (yellow) are shown.}
\label{fig:3dbands}
\end{figure}

 The calculated PBE and HSE06 electronic band gaps $E_g$ are shown in Table \ref{tab:tabela-gap}. As can be seen, the PBE bandgaps lie in the range of 2.17-3.47 eV. However, it is worth mentioning that these values are not accurate when compared with experimental data because of the well-known band gap underestimation associated with self-interaction errors \cite{Tsuneda,SIE-DFT,DUAN201556,Lentz_2020}. Despite this underestimation, it is noteworthy that some PBE results already fall within the wide-bandgap semiconductor regime, highlighting the intrinsically large band gaps of these materials. A more accurate description is provided by the hybrid HSE06 functional, which partially corrects for the self-interaction error by including a portion of exact exchange \cite{hse}. Within this framework, the predicted band gaps range from 3.56 to 5.25 eV, placing all systems in the category of (UWBG) semiconductors \cite{uwbg1,uwbg2}, typically defined as those with $E_g > 3.4$ eV. Also, as previously discussed, the nature of the band gap varies among the different phases of 2D-\ce{GeO2}, with some exhibiting indirect gaps and others showing direct gaps, which could further influence their suitability for specific device implementations.
 
\begin{table}[h!]
\centering
\caption{Comparison between PBE and HSE06  electronic band gaps $E_g$ of 2D-\ce{GeO2}.}
\begin{tabular}{l|cc|cc}
\hline
\multicolumn{1}{c|}{} & \multicolumn{2}{c|}{DFT-HSE06} & \multicolumn{2}{c}{DFT-PBE} \\ \hline
\multicolumn{1}{c|}{Space group} & $E_g$(eV)  & bandgap type & $E_g$(eV)  & bandgap type\\ \hline
P-3m1  & 5.20  & Indirect & 3.47 & Indirect \\
P6/mmm & 5.25 & Direct & 3.41 & Direct \\
Pbcm   &   4.92  & Direct & 3.03 & Direct \\
Pmma   &   4.39  & Direct & 2.45 & Direct \\
P4m2(1) & 4.86 & Direct & 2.93 & Direct \\
P4m2(2) & 4.77 & Indirect & 2.92 & Indirect \\
Pmmm(1) & 4.68 & Indirect & 2.93 & Indirect \\
Pmmm(2) & 3.56 & Indirect & 2.17 & Indirect \\ \hline
    \end{tabular}
    \label{tab:tabela-gap}
\end{table}

\begin{figure}[H]
\centering
\includegraphics[width=0.9\linewidth]{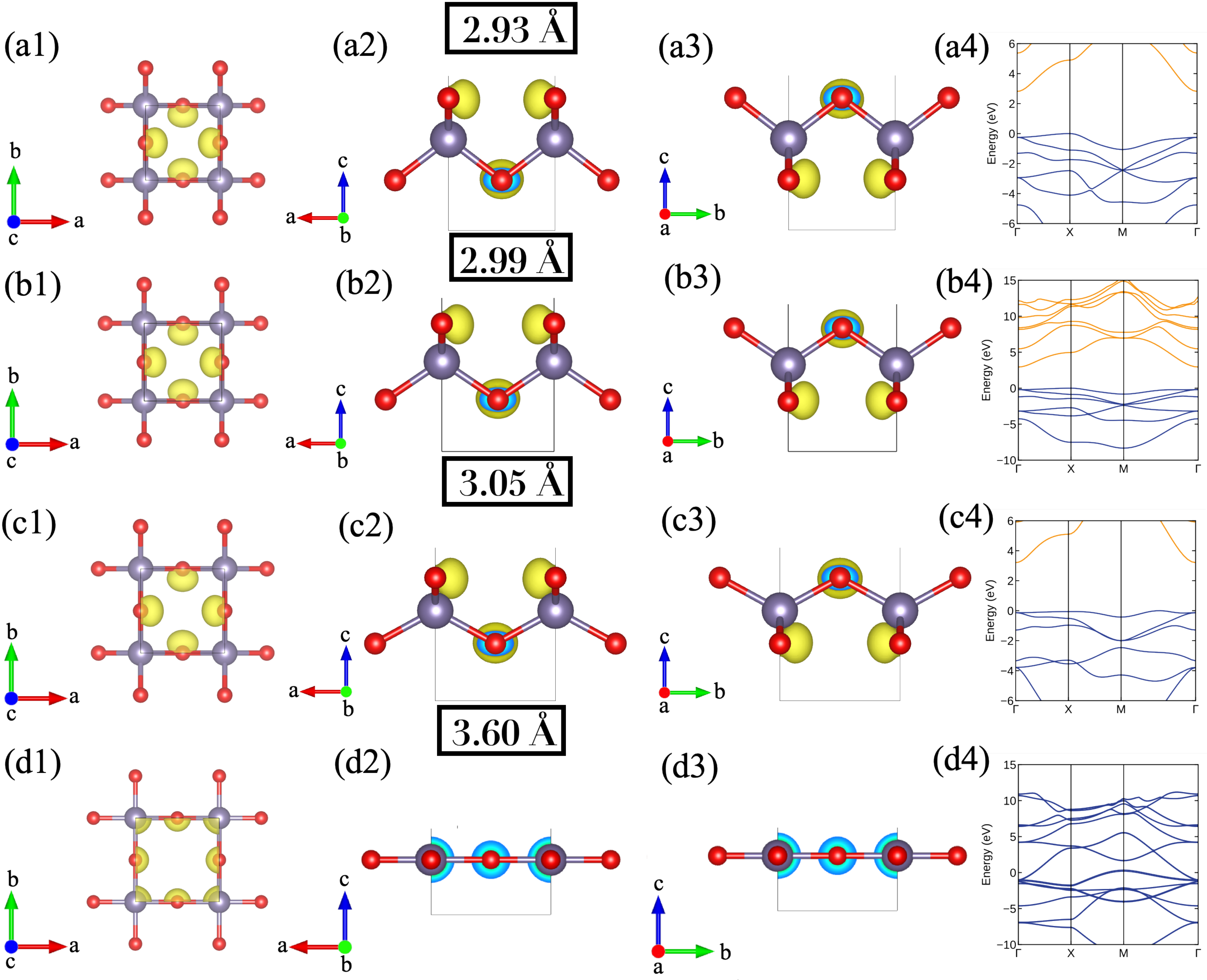}
\caption{Projected charge densities for VBM and electronic band structures of 2D \ce{GeO2} in the P4m2(2) phase for several lattice parameters: (a) $a$ = 2.93\,{\AA}, (b) $a$ = 2.99\,{\AA}, (c) $a$ = 3.05\,{\AA} and (d) $a$ = 3.60\,{\AA}.}
\label{fig:plotcharge}
\end{figure}

In order to investigate the appearance and tunability of flat bands in the valence region, we use the orthorhombic P4m2(2) phase of the two-dimensional \ce{GeO2} monolayer as a showcase. In this geometry, the valence band structure exhibits pronounced flat bands near the valence band maximum (VBM). These features are a hallmark of spatially localized electronic states, which arise due to minimal orbital hybridization and weak interatomic overlap.

Orbital-resolved band structures shown in Fig. \ref{fig:pbands} and orbital resolved charge density shown in Fig.\,\ref{fig:plotcharge} reveal that these flat states originate predominantly from non-bonding O-$p$ orbitals, particularly the in-plane $p_{\rm x}$ and $p_{\rm y}$ components. The charge density is strongly localized around oxygen atoms, consistent with the presence of dangling bonds due to undercoordination. Each oxygen atom in the monolayer is bonded to three neighboring Ge atoms, leaving one p$_{\rm xy}$ orbital unbonded. This orbital localization gives rise to the flat dispersion observed near the VBM.

A key structural feature that modulates this behavior is the degree of buckling in the Ge–O–Ge bridges. In coplanar configurations, where Ge and O atoms lie in the same plane, the orbital alignment enhances Ge–O hybridization, resulting in broader, more delocalized valence bands. In contrast, buckled structures reduce the effective overlap between Ge sp$^3$-like orbitals and O-$p$ orbitals, preserving the non-bonding character of the oxygen states and sustaining the flat-band character.

To illustrate this behavior, we calculated the projected charge density for the P4m2(2) phase of \ce{GeO2} under varying in-plane lattice parameters. Figs.~\ref{fig:plotcharge}(a1)-(a3) and Figs.\ref{fig:plotcharge}(c1)-(c3) correspond to compressive ($a = 2.93$~\AA) and tensile ($a = 3.05$~\AA) strain conditions, respectively. For completeness, the unstrained ground-state structure ($a = 2.99$~\AA) is shown in Fig.~\ref{fig:plotcharge} (b1)-(b3). In all cases, the charge densities of the VBM remains highly localized on the oxygen atoms, and the associated O-p bands retain their flat character. The corresponding band structures, presented in Fig.~\ref{fig:plotcharge}(a4)-(c4), further confirm that under moderate biaxial strain ($\pm$2\%), the flat-band nature of the VBM is preserved with only minor modifications in dispersion. This reveals that flat bands in 2D \ce{GeO2} can be tuned via strain without significantly disrupting their localized orbital origin. In addition, we analyzed the behavior of the charge density and the electronic band structure for a large strain applied $\sim$20.4 $\%$ ($a = 3.60$~\AA). In such case, the orbital alignment is enhanced, allowing for stronger hybridization between the Ge and O atoms [Fig.~\ref{fig:plotcharge} (d1)-(d3)]. This increased overlap leads to delocalized states and the disappearance of the flat bands in the valence region [Fig.~\ref{fig:plotcharge} (d4)]. Thus, the flat valence bands in monolayer \ce{GeO2} emerge from a combination of localized, non-bonding O $\rm p$ orbitals and structural factors—particularly buckling-induced suppression of Ge–O hybridization.

Owing to this localization the Chern and Z$_2$ invariants\cite{WU2017}
for all germanium dioxide monolayers are calculated. It turns out that
the integration along the entire Brillouin zone yields null values for
both topological invariants. This result indicates the absence of
nontrivial topological phases in the examined \ce{GeO2} polymorphs,
confirming that they are all topologically trivial insulators. We
believe that perhaps the absence of strong spin-orbit
coupling in GeO$_2$ is responsible for the absence of quantum anomalous Hall states.
phases.

To further explore the optoelectronic response of the system, particularly its excitonic behavior, we turn our attention to its optical properties. Given the high computational cost associated with incorporating excitonic effects via the DFT formalism as implemented in VASP, we instead employ the Bethe–Salpeter equation (BSE) within the tight-binding (TB) framework - a robust and time-effective approach for calculating excitonic spectra. The orthogonal TB Hamiltonian, $H(\mathbf{k})$, obtained within the Wannier90 framework \cite{wannier90,PhysRevB74-195118} as described in SI.

In 2D materials, due the quantum confinement in their non-periodic
direction, excitonic quasi-particle effects are significant. The exciton band structure
of monolayered \ce{GeO2} phases calculated by solving the BSE on top
of MLWF-TB-HSE06 calculations along the electronic band structure
\textbf{k}-path is depicted in Fig.\,\ref{fig:exc_bands}. The exciton
band structures enable the identification of whether the excitonic
ground state ($Ex_{gs}$) is direct or indirect.  The exciton ground
state is indirect regardless of the crystal phase. This means that the
electron and hole single particle states are located at different
\textbf{k}-points. Indeed, this is for P-3m1, P4m2(2), Pmmm(1), and
Pmmm(2) phases as those systems have an indirect band gap. However,
phases with direct band gaps the opposite behavior is expected. A
possible explanation could be the character of the electron and hole
orbital symmetries. A similar behavior was previously found for 1T'
\ce{RuO2}\cite{Santos_3677_2023} and 1T'
\ce{OsO2}\cite{Santos_074301_2023} monolayers.

\begin{figure}[H]
    \centering
    \includegraphics[width=0.9\linewidth]{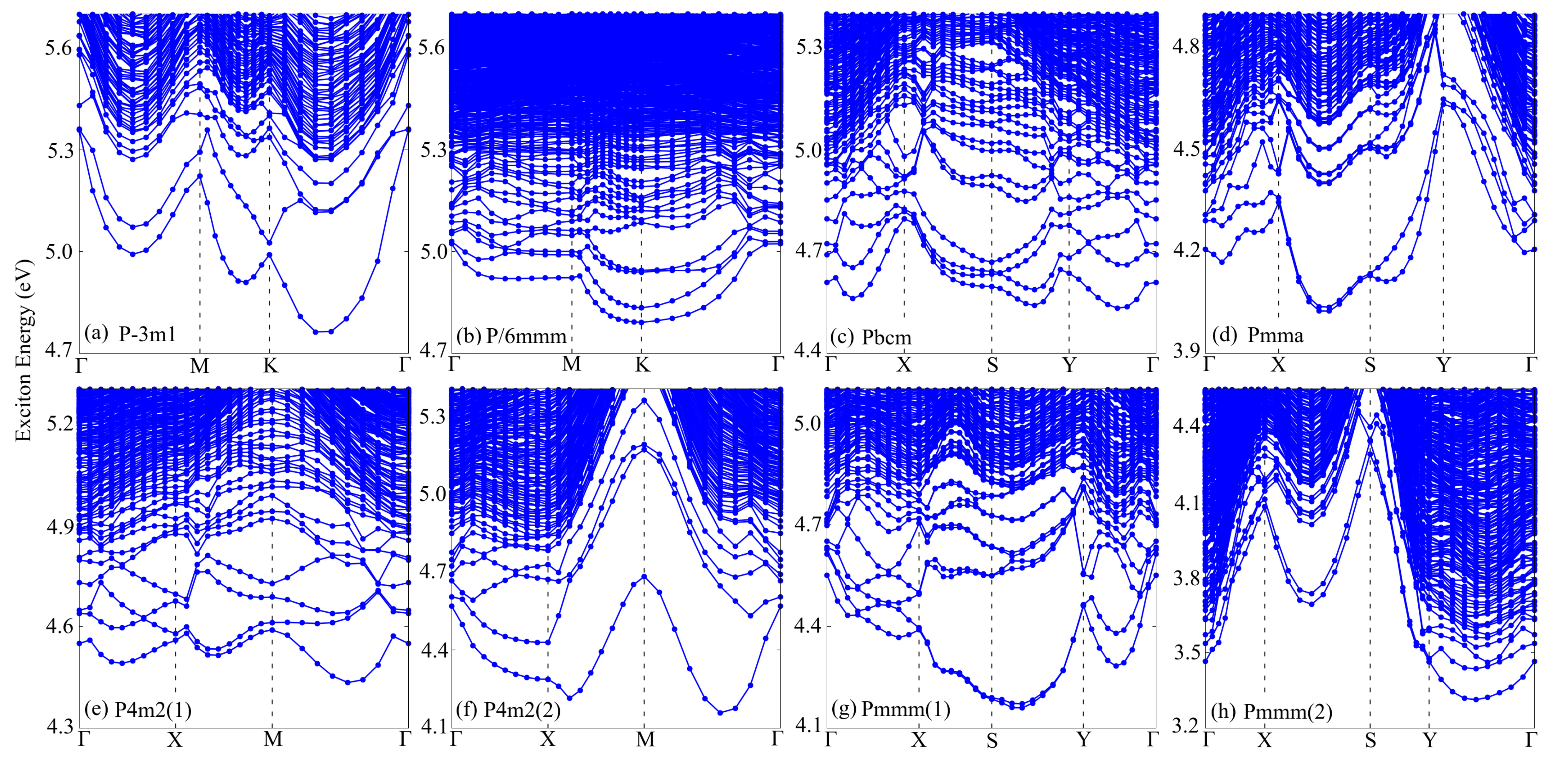}
    \caption{Exciton band structure of 2D-\ce{GeO2}: a) P-3m1, b) P6/mmm, c) Pbcm, d) Pmma, e) P4m2(1), f) P4m2(2), g) Pmmm(1) and h) Pmmm(2).}
    \label{fig:exc_bands}
\end{figure}

The exciton binding energy $Ex_{b}$ of germanium dioxide monolayers are estimated from the difference between electronic fundamental band gap $E_{g}$ and exciton ground state $Ex_{gs}$, as shown in Table~\ref{tab:exciton_data}. The excitonic effects are strong, with binding energies lying in the range of \SIrange{194}{608.32}{\milli\electronvolt}. This is a similar to what has been found in other 2D materials,\cite{Dias_3265_2021,Moujaes_111573_2023,Guassi_11710_2024,Dias_8572_2024}. The larger $Ex_{b}$ appears for the P4m2(2) phase  and the smaller value for the Pmmm(2) phase, showing that the exciton binding energy is highly dependent of crystal symmetry.  Previous BSE-${\rm G_0W_0}$ calculations \cite{D1CP02299G} for the P-3m1 phase reported an exciton binding energy of \SI{1.55}{\electronvolt}, while our results  BSE calculations shows \SI{0.44}{\electronvolt}. This difference could be justified first for the much larger G$_0$W$_0$ electronic band gap of \SI{7.20}{\electronvolt} compared to  the HSE06 band gap of \SI{5.20}{\electronvolt}.  For a system with a large gap, the screening is weak and the exciton binds strongly, giving rise to a relatively narrow
spatial extension and high binding energy. On the other hand, for a system with a small gap, the screening is strong and the exciton binding energy smaller. It is thus expected a larger exciton binding energy for larger electronic band gaps,\cite{Choi_066403_2015,Zhang_209701_2017} although this relationship is not always linear\cite{PhysRevLett.118.266401,PhysRevLett.115.066403}. Another cause of discrepancy might be the \textbf{k}-mesh used for BSE simulations, as the mentioned works used a $18\times18\times1$ and our calculations used $47\times47\times1$, excitonic effects are very dependent of \textbf{k}-mesh density for accurate results. As a matter of comparison, The exciton binding energy in boron nitride (BN) varies depending on its dimensionality, ranging from approximately \(0.7\) eV in bulk hexagonal BN to around \(2.1\) eV in a single monolayer\,\cite{PhysRevLett.96.126104} 

\begin{table}[h!]
\centering
\caption{Excitonic properties obtained by MLWF-TB+BSE at HSE06:  fundamental band gap ($E_{g}$), direct band gap  ($E^{d}_{g}$), exciton ground state, ($Ex_{gs}$) direct exciton ground state ($Ex^{d}_{gs}$). The exciton binding energy ($Ex_{b}$) is  calculated as $E_{g}-Ex_{gs}$.}
\begin{tabular}{lccccc} \toprule
Space group      & $E_{g}$ (\si{\electronvolt}) & $E^{d}_{g}$ (\si{\electronvolt}) & $Ex_{gs}$ (\si{\electronvolt}) & $Ex^{d}_{gs}$ (\si{\electronvolt})  & $Ex_{b}$ (\si{\milli\electronvolt}) \\ \midrule
P-3m1 & 5.20  & 5.67  &4.76  &5.36  &437.54 \\ 
P6/mmm &5.25  &5.25  &4.79  &5.02  &464.47 \\
Pbcm &4.92  &4.92  &4.53  &4.61  &386.81 \\ 
Pmma &4.39  &4.39  &4.02  &4.20  &367.70 \\ 
P4m2(1) &4.86  &4.86  &4.43  &4.55  &424.33 \\
P4m2(2) &4.77  &4.89  &4.16  &4.57  &608.32 \\ 
Pmmm(1) &4.68  &4.79  &4.16  &4.55  &519.13 \\ 
Pmmm(2) &3.56  &3.63  &3.36  &3.51  &194.31 \\
 \bottomrule
    \end{tabular}
    \label{tab:exciton_data}
\end{table}

The linear optical response of the \ce{GeO2} monolayers are investigated by calculating the absorption spectrum, as shown in Fig.~\ref{fig:exc_optics}. Due to the significant exciton binding energies of these monolayers,  this  results in a red shift of the optical band gap between \SIrange{200}{600}{\milli\electronvolt}. All systems absorbs in UV region. In addition, the Pmmm(2) phase also shows absorption in the visible region. The  P6/mmm, P4m2(1) and P4m2(2) phases show an isotropic behavior at IPA level of calculation. As quasi-particle effects are taken in account, a very small optical anisotropy is seen. This sort of behavior was previously reported for other 2D materials\,\cite{Moujaes_111573_2023,Dias_8572_2024,Huacarpuma_6634_2025}.

The P-3m1 phase, Fig. \ref{fig:exc_optics} shows an optical anisotropy at IPA and BSE levels. However, it shows identical peaks, with different intensities (higher along the $\hat{x}$-polarized light). On the contrary, the Pbcm, Pmma and Pmmm(2) phases shows, for calculations at BSE and IPA levels, distinct absorption peaks. Moreover, the optical band gaps due to incident light polarization is only dipole-allowed for $\hat{x}$-polarization, resulting in a blue shift of optical band the gap for $\hat{y}$ polarized light.

As a general feature, the absorption coefficient values significantly change with the direction of the incident polarized light. Pmmm(1) also shows optical anisotropy, but this has a negligible effect on the optical band gap. 

Surprisingly, at photon excitation energies around \SIrange{6}{6.7}{\electronvolt}, the BSE absorption coefficients are significantly larger for polarization along $\hat{x}$, and the opposite happens at the IPA level.  The underlying mechanism for this is still not clear, but one can speculate that the electron and hole orbital symmetries and also the electron-hole Coulomb interaction play a role. One cannot rule out the role of the electron-phonon interaction, and further investigations are needed to address this point.

\begin{figure}[H]
    \centering
    \includegraphics[width=0.9\linewidth]{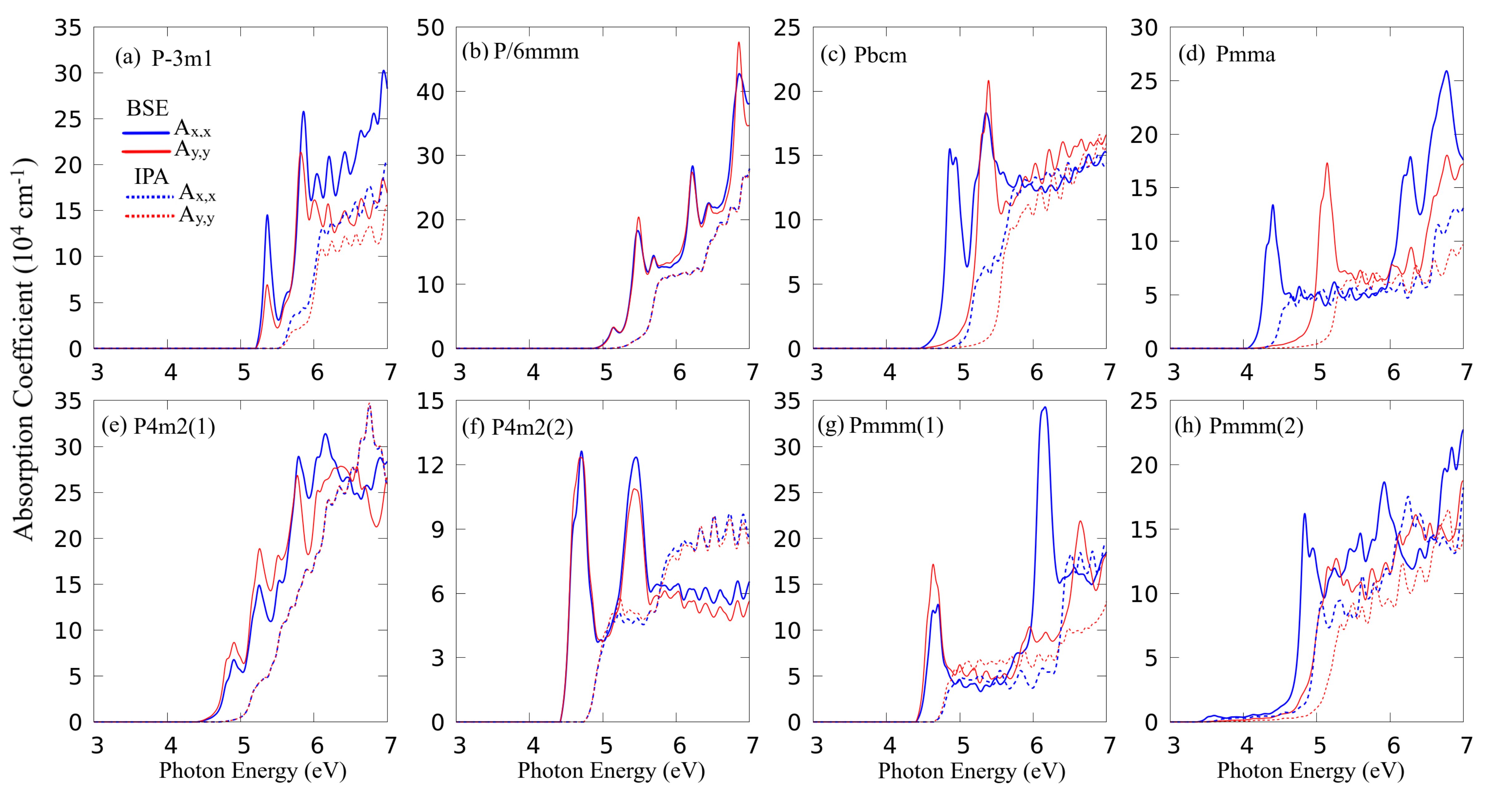}
    \caption{Absorption coefficient of germanium dioxide monolayers at BSE (solid curves) and IPA (dashed curves) levels.  Linear light polarization along $\hat{x}$ (blue curves) and $\hat{y}$ (red curves) directions are considered. a) P-3m1, b) P6/mmm, c) Pbcm, d) Pmma, e) P4m2(1), f) P4m2(2), g) Pmmm(1) and h) Pmmm(2).}
    \label{fig:exc_optics}
\end{figure}

As a matter of completeness we have also calculated refractive index
and the reflectibility at IPA and BSE levels as shown in
Figs. S1-S8. Finally we have included the substrate in the calculations of the absorption coefficent by using the experimental static dielectric constant of lattice-matched substrates to the P6/mmm GeO$_2$ phase. To mimic the substrates the monolayered WTe$_2$ is used. The employed dielectric constant beneath the oxide is $\epsilon$ = 6.19 for WSe$_2$\cite{epsilon_WSe2}. Negligible difference is seen compared to the calculations done of  the absorption coefficient layers calculated when a vacuum dielectric constant is used beneath the GeO$_2$ layer, as shown in Fig.Sxx.

\section{Conclusions}

Our first-principles and MLWF-BSE calculations demonstrates that all 2D
GeO$_2$ polymorphs are ultra-wide-band-gap semiconductors
with robust excitonic effects. The valence bands,
predominantly composed of localized O-p orbitals, are nearly flat
across the Brillouin zone, resulting in strong carrier localization
that is preserved under moderate in-plane strain. These flat-band
features can be modulated through strain or structural buckling,
enabling control over electronic dispersion without compromising
thermal stability.

No nontrivial topological phases are identified, confirming that all
2D-GeO$_2$ polymorphs are topologically trivial insulators. Nonetheless,
the coexistence of tunable ultra-wide band gaps, strong excitonic
coupling, and flat-band localization opens avenues for next-generation
oxide electronics, deep-UV photonics, and potentially correlated
states in doped or heterostructured GeO$_2$ layers. These results
position monolayer GeO$_2$ as a versatile building block for power
electronics and optoelectronic platforms operating under extreme
conditions.

\section{Acknowledgments}

All authors acknowledge the financial support from the Brazilian
funding agency CNPq (Conselho Nacional de Desenvolvimento
Cient\'{\i}fico e Tecnol\'ogico) grant numbers 305174/2023-1,
313081/2017-4, 305335/2020-0, 309599/2021-0, 408144/2022-0,
305952/2023-4, 444069/2024-0 and 444431/2024-1. G.E.G.A. thanks for a
CNPq scholarship project number PI05133-2020. A.C.D acknowledge FAPDF
grant numbers 00193-00001817/2023-43 and 00193-00002073/2023-84. A.C.D
and A.L.R acknowledge FAPDF-CAPES grant number
00193-00000867/2024-94. A. L. R. thanks Mauricio Chagas da Silva for
fruitful discussions.
All authors thank computational resources from
LaMCAD/UFG, Santos Dumont/LNCC, CENAPAD-SP/Unicamp (project number 897
and 761) and Lobo Carneiro High Performance Computer (project number
133).


\providecommand{\latin}[1]{#1}
\makeatletter
\providecommand{\doi}
  {\begingroup\let\do\@makeother\dospecials
  \catcode`\{=1 \catcode`\}=2 \doi@aux}
\providecommand{\doi@aux}[1]{\endgroup\texttt{#1}}
\makeatother
\providecommand*\mcitethebibliography{\thebibliography}
\csname @ifundefined\endcsname{endmcitethebibliography}
  {\let\endmcitethebibliography\endthebibliography}{}

\clearpage
\newpage

\begin{center}
{\bf Supporting Information:} \\ 
{\bf Flat bands in ultra-wide gap two-dimensional germanium dioxide}
\end{center}

\section{Computational Details}

\begin{table}[H]
\centering
\caption{Essential information for the chosen PAW projectors (\texttt{POTCAR}), which encompasses key details such as the PAW-PBE projector name, the creation date, the count of valence electrons ($Z_{val}$) and the maximum recommended cutoff energy (\texttt{ENMAX}), for each of the selected chemical species.}
\begin{tabular}{llccc} \toprule
Element & \texttt{POTCAR} & Date            & $Z_{val}$         & \texttt{ENMAX}       \\
        & PAW-PBE         & \texttt{POTCAR} &           &                 (\si{\electronvolt}) \\ \midrule
\ce{Ge}  &\ce{Ga}     & 05/01/2001       &4           &173.807               \\
\ce{O}  &\ce{O}    & 08/04/2002       &6            &400.000               \\  \bottomrule
\end{tabular}
\end{table}

\subsection{POSCAR Optimized Structures}

\subsubsection{P-3m1}

\texttt{P-3m1 \\                          
   1.00000000000000   \\     
     2.9186048000000002    0.0000000000000000    0.0000000000000000 \\  
    -1.4593024000000001    2.5275859000000001    0.0000000000000000 \\  
     0.0000000000000000    0.0000000000000000   25.0000000000000000 \\  
   Ge   O  \\  
     1     2 \\  
Direct \\  
  0.0000000000000000  0.0000000000000000  0.5000000000000000 \\  
  0.3333356114478505  0.6666643885521495  0.5392990000758218 \\  
  0.6666643885521495  0.3333356114478505  0.4607009999241782}

\subsubsection{P/6mmm}

\texttt{P6/mmm \\                            
   1.00000000000000 \\       
     5.4976487479999996    0.0000000000000000    0.0000000000000000 \\  
    -2.7488243739999998    4.7611034768499998    0.0000000000000000 \\  
     0.0000000000000000    0.0000000000000000   25.0000000000000000 \\  
   Ge   O  \\  
     4     8 \\  
Direct \\  
  0.6819530901089834  0.3288561483535872  0.4293448561513031 \\  
  0.3480964393592743  0.6612536793792358  0.5706694329446123 \\  
  0.3480893917643400  0.6608064746206210  0.4293287473231615 \\  
  0.6807660860599043  0.3278217661493912  0.5706527802581860 \\  
  0.4191012688633364  0.3989703899110992  0.4038172634490422 \\  
  0.6111388098145554  0.5915062301164653  0.5959361776524332 \\  
  0.6148019194216587  0.9950417892312302  0.4050225062674002 \\  
  0.4198056727973452  0.9944836190553303  0.5962821934665499 \\  
  0.0151184505140876  0.5896492361738339  0.4036655444927177 \\  
  0.0140606476945351  0.3955101580656049  0.5952874319616015 \\  
  0.3319847200059129  0.6680353959017182  0.4999967642366556 \\  
  0.6650834735960700  0.3380650530418876  0.4999963617963417}

\subsubsection{Pbcm}

\texttt{Pbcm \\                             
   1.00000000000000 \\       
     5.9184871077365475    0.0000000000000000    0.0000000000000000 \\  
     0.0000000000000000    5.2292344165842906    0.0000000000000000 \\  
     0.0000000000000000    0.0000000000000000   16.0000000000000000 \\  
   Ge   O  \\  
     4     8 \\  
Direct \\  
  0.2500004085579235  0.6845286680623275  0.3384777360953777 \\  
  0.7499995336236225  0.3154709526478356  0.2153035139046224 \\  
  0.2500004085579235  0.1845298223712248  0.2153035139046224 \\  
  0.7499995336236225  0.8154717276668961  0.3384777360953777 \\  
  0.2500004085579235  0.3502669746621834  0.3137224920710722 \\  
  0.7499995336236225  0.6497326460479798  0.2400587579289279 \\  
  0.2500004085579235  0.8502658203532860  0.2400587579289279 \\  
  0.7499995336236225  0.1497338003568771  0.3137224920710722 \\  
  0.5000008171158470  0.7499992332006329  0.3986490784223378 \\  
  0.5000008171158470  0.2500003875095302  0.1551327965776679 \\  
  0.0000000000000000  0.2500003875095302  0.1551327965776679 \\  
  0.0000000000000000  0.7499992332006329  0.3986490784223378}

\subsubsection{Pmma}

\texttt{Pmma \\                             
   1.00000000000000 \\       
     3.8138524576652855    0.0000000000000000    0.0000000000000000 \\  
     0.0000000000000000    3.0551484114956575    0.0000000000000000 \\  
     0.0000000000000000    0.0000000000000000   23.2759840228455701 \\  
   Ge   O  \\  
     2     4 \\  
Direct \\  
  0.2500000000000000  0.0000000000000000  0.5503387769063082 \\  
  0.7500000000000000  0.0000000000000000  0.4496612230936918 \\  
  0.2500000000000000  0.0000000000000000  0.4693350424508509 \\  
  0.7500000000000000  0.0000000000000000  0.5306649275491537 \\  
  0.7500000000000000  0.5000000000000000  0.4087546312681170 \\  
  0.2500000000000000  0.5000000000000000  0.5912453387318877}

\subsubsection{P4m2(1)}

\texttt{P4m2(1) \\                                  
   1.00000000000000  \\      
     5.9816945617522146    0.0000000000000000    0.0000000000000000 \\  
     0.0000000000000000    5.9816945617522146    0.0000000000000000 \\  
     0.0000000000000000    0.0000000000000000   16.0000000000000000 \\  
   Ge   O  \\  
     4     8 \\  
Direct \\  
  0.4999996315390405  0.7475094597061300  0.4733194100742040 \\  
  0.4999996315390405  0.2524898033719580  0.4733194100742040 \\  
  0.7475094597061300  0.4999996315390405  0.3369787149257988 \\  
  0.2524898033719580  0.4999996315390405  0.3369787149257988 \\  
  0.4999996315390405  0.4999996315390405  0.2713338658317070 \\  
  0.4999996315390405  0.4999996315390405  0.5389642591682886 \\  
  0.4999996315390405  0.0000000000000000  0.5333834208070698 \\  
  0.0000000000000000  0.4999996315390405  0.2769147041929330 \\  
  0.7409795898676848  0.2590196732103962  0.4051493750000006 \\  
  0.2590196732103962  0.7409795898676848  0.4051493750000006 \\  
  0.2590196732103962  0.2590196732103962  0.4051493750000006 \\  
  0.7409795898676848  0.7409795898676848  0.4051493750000006}

\subsubsection{P4m2(2)}

\texttt{P4m2(2) \\                             
   1.00000000000000  \\      
     3.0133888824762951    0.0000000000000000    0.0000000000000000 \\  
     0.0000000000000000    3.0133888824762951    0.0000000000000000 \\  
     0.0000000000000000    0.0000000000000000   16.0000000000000000 \\  
   Ge   O  \\  
     1     2 \\  
Direct \\  
  0.0000000000000000  0.0000000000000000  0.1498893750000008 \\  
  0.4999992415986938  0.0000000000000000  0.0872462066238668 \\  
  0.0000000000000000  0.4999992415986938  0.2125325433761347}

\subsubsection{Pmmm(1)}

\texttt{Pmmm(1) \\                             
   1.00000000000000  \\      
     2.9107279409552809    0.0000000000000000    0.0000000000000000 \\  
     0.0000000000000000    3.6817194594027320    0.0000000000000000 \\  
     0.0000000000000000    0.0000000000000000   19.3841784134130037 \\  
   Ge   O  \\  
     2     4 \\  
Direct \\  
  0.5000000000000000  0.0000000000000000  0.5092076405130257 \\  
  0.0000000000000000  0.5000000000000000  0.3927921760475428 \\  
  0.0000000000000000  0.0000000000000000  0.5643286809347217 \\  
  0.0000000000000000  0.0000000000000000  0.4269442801918544 \\  
  0.5000000000000000  0.5000000000000000  0.3376717879578024 \\  
  0.5000000000000000  0.5000000000000000  0.4750554943550682}

\subsubsection{Pmmm(2)}

\texttt{Pmmm(2)                           
   1.00000000000000  \\      
     5.0777623020657003    0.0000000000000000    0.0000000000000000 \\  
     0.0000000000000000    2.8641663690431653    0.0000000000000000 \\  
     0.0000000000000000    0.0000000000000000   28.7464348837111245 \\  
   Ge   O  \\  
     2     4 \\  
Direct \\  
  0.5000000000000000  0.0000000000000000  0.5181758745611802 \\  
  0.0000000000000000  0.5000000000000000  0.4854915532396475 \\  
  0.2637081179068446  0.5000000000000000  0.5376426064729571 \\  
  0.7637073094359721  0.0000000000000000  0.4660236831266360 \\  
  0.7362919120931508  0.5000000000000000  0.5376426064729571 \\  
  0.2362926755640231  0.0000000000000000  0.4660236831266360}

\subsection{Structure relaxation}

\noindent Exchange-correlation functionsl (PBE)\\
\noindent ENCUT = 450 eV\\
\noindent EDIFF = 1E-6 eV\\
\noindent EDIFFG = -0.01 eV/{\AA}\\
\noindent K-points: 6x6x1\\
\noindent Vacuum region: 15 {\AA}\\

\subsection{AIMD calculations}

\begin{table}[H]
\centering
\caption{Calculation parameters for AIMD. The structures were brought in contact with a Nosé-Hoover thermostat at T = 300K. Simulation time was set to 10ps with time step of 0.5 fs. k-point sampling at $\gamma$ point.}
\begin{tabular}{lcc} \toprule
symmetry & size &  numer of atoms\\
\hline
P-3m1 & 5x5x1 & 75 \\
P6/mmm & 2x2x1 & 48 \\
Pbcm & 1x2x2 & 48 \\
Pmma & 3x3x1 & 54 \\
P4m2(1) & 2x2x1 & 48\\ 
P4m2(2) & 4x4x1 & 48 \\
Pmmm(1) & 4x4x1 & 96 \\
Pmmm(2) & 4x4x1 & 96 \\
\bottomrule
\end{tabular}
\end{table}

\begin{figure}[h]
    \centering
    \includegraphics[width=0.9\linewidth]{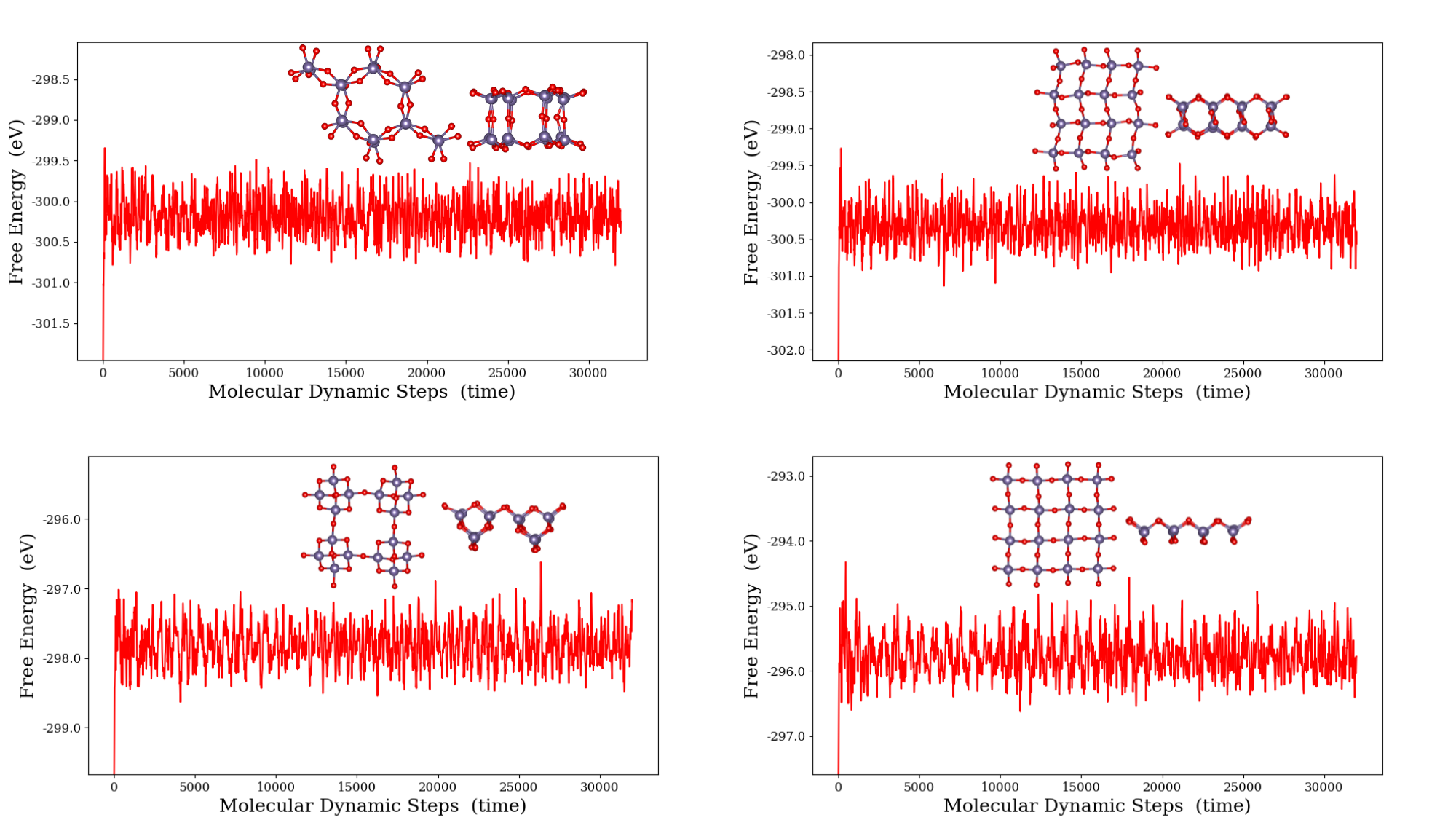}
    \caption{Free energy profiles from AIMD simulations of selected structures for simulation time of 16 ps at 300 K. The increasing of the simulation time did not modify the previous achievements (Fig. S1). The insets represent the corresponding atomic configurations at the end of simulation.}
    \label{AIMD}
\end{figure}

\subsection{Phonons calculations}

\begin{table}[H]
\centering
\caption{Phonon calculations for germanium dioxide. Input generation and interpolation with Phonopy calculated with the finite displacement method. Energy cutoff set to 520 eV with energy convergence criterium of 10$^{-8}$ eV and forces on atoms of 10$^{-3}$ eV/{\AA}. k-point sampling at gamma point.}
\begin{tabular}{lcc} \toprule
symmetry & supercell size & atoms\\
\hline
P-3m1 & 6x6x1 & 108  \\
P6/mmm & 3x3x1 & 108 \\
Pbcm & 3x3x1 & 108 \\
Pmma & 4x4x1 & 96 \\
P4m2(1) & 3x3x1 & 108  \\
P4m2(2) & 6x6x1 & 108  \\
Pmmm(1) & 4x4x1 & 96 \\
Pmmm(2)& 4x4x1 & 96 \\
\bottomrule
\end{tabular}
\end{table}

\section{Excitonic and Optical Properties}

\subsection{BSE Simulation parameters}

\begin{table}[H]
\centering
\caption{Parameters used for BSE simulations:\textbf{k}-points density , $R_{k}$ (\si{\per\angstrom}) and their correspondent \textbf{k}-mesh, $n_v$, number of valence bands, $n_c$, number of conduction bands and dielectric function smearing $\eta$ (\si{\electronvolt}). All simulations were done using the Coulomb truncated 2D potential (V2DT) with the system surrounded by vacuum, implemented in WanTiBEXOS package.}

    \begin{tabular}{lcccccc} \toprule
    Space group  & $R_k$ & \textbf{k}-mesh   & $n_c$ & $n_v$ & $\eta$ \\ \midrule

    P-3m1    &120  &$47\times47\times1$   &1  &5   & 0.05\\
    P6/mmm   &120  &$25\times25\times1$   &2  &13  & 0.05\\
    Pbcm     &120  &$20\times23\times1$   &2  &12  & 0.05\\  
    Pmma     &120  &$31\times39\times1$   &2  &6   & 0.05\\
    P4m2(1)  &120  &$20\times20\times1$   &2  &11  & 0.05\\
    P4m2(2)  &120  &$40\times40\times1$   &1  &5   & 0.05\\
    Pmmm(1)  &120  &$41\times33\times1$   &2  &7   & 0.05\\ 
    Pmmm(2)  &120  &$24\times42\times1$   &3  &8  & 0.05 \\
     \bottomrule
    \end{tabular}
    \label{tab:bse_params}
\end{table}

\subsection{Optical Properties}

\begin{figure}[H]
    \centering
    \includegraphics[width=0.8\linewidth]{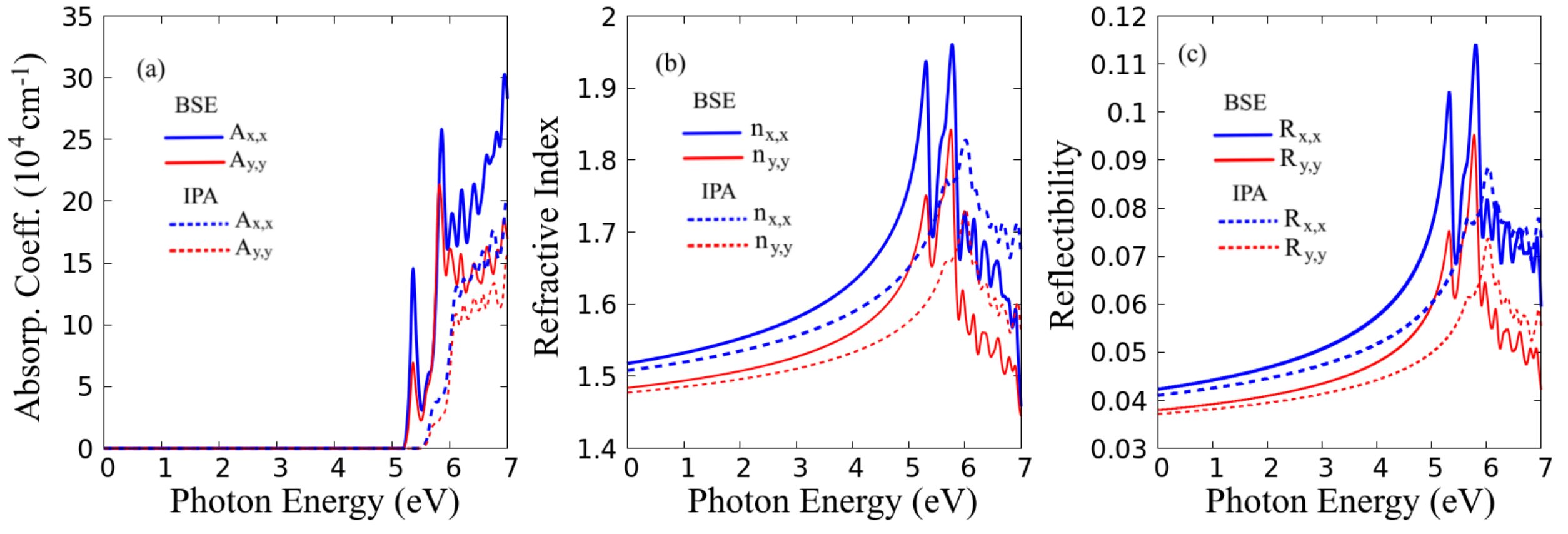}
    \caption{Optical properties of P-3m1 phase of \ce{GeO2} monolayer, (a) refractive index and (b) reflectibility, at IPA (dashed curves) and BSE (solid curves) levels, considering a linear light polarization along $\hat{x}$ (blue curves) and $\hat{y}$ (red curves) directions.}
    \label{fig:opt_p-3m1}
\end{figure}

\begin{figure}[H]
    \centering
    \includegraphics[width=0.8\linewidth]{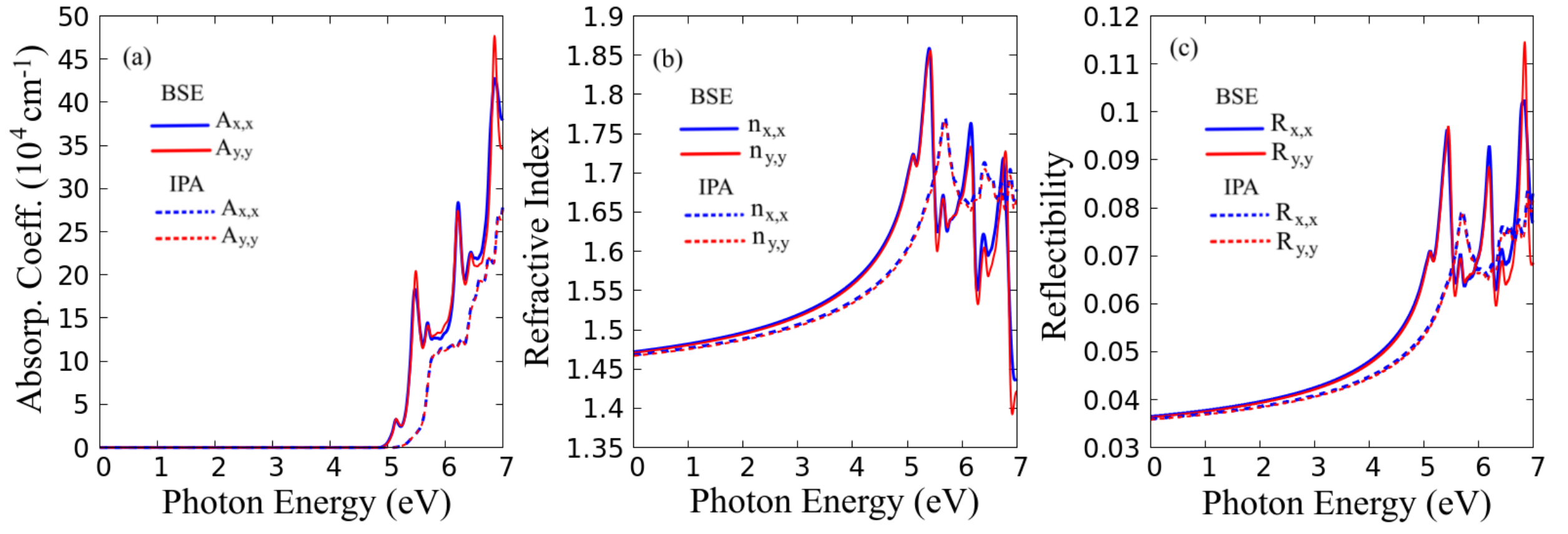}
    \caption{Optical properties of P/6mmm phase of \ce{GeO2} monolayer, (a) refractive index and (b) reflectibility, at IPA (dashed curves) and BSE (solid curves) levels, considering a linear light polarization along $\hat{x}$ (blue curves) and $\hat{y}$ (red curves) directions.}
    \label{fig:opt_P/6mmm}
\end{figure}

\begin{figure}[H]
    \centering
    \includegraphics[width=0.8\linewidth]{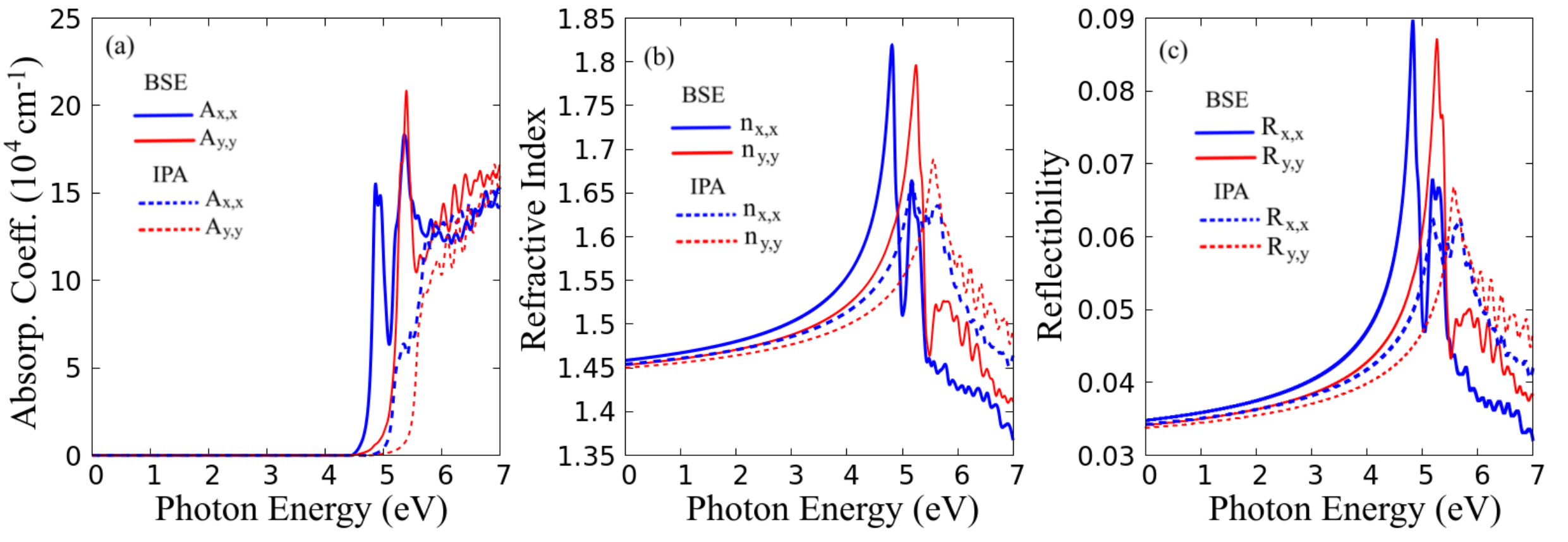}
    \caption{Optical properties of Pbcm phase of \ce{GeO2} monolayer, (a) refractive index and (b) reflectibility, at IPA (dashed curves) and BSE (solid curves) levels. Linear light polarization along  $\hat{x}$ (blue curves) and $\hat{y}$ (red curves) directions was considered.}
    \label{fig:opt_pbcm}
\end{figure}

\begin{figure}[H]
    \centering
    \includegraphics[width=0.8\linewidth]{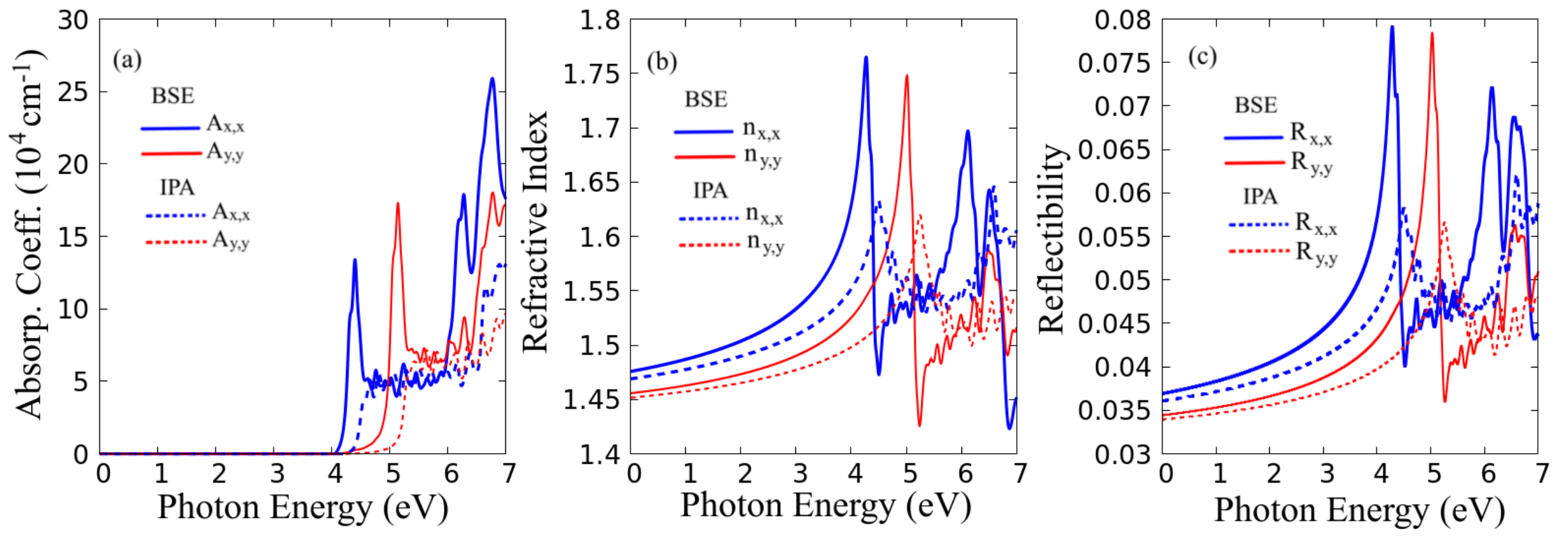}
    \caption{Optical properties of Pmma phase of \ce{GeO2} monolayer, (a) refractive index and (b) reflectibility, at IPA (dashed curves) and BSE (solid curves) levels, considering a linear light polarization at $\hat{x}$ (blue curves) and $\hat{y}$ (red curves) directions.}
    \label{fig:opt_pmma}
\end{figure}   

\begin{figure}[H]
    \centering
    \includegraphics[width=0.8\linewidth]{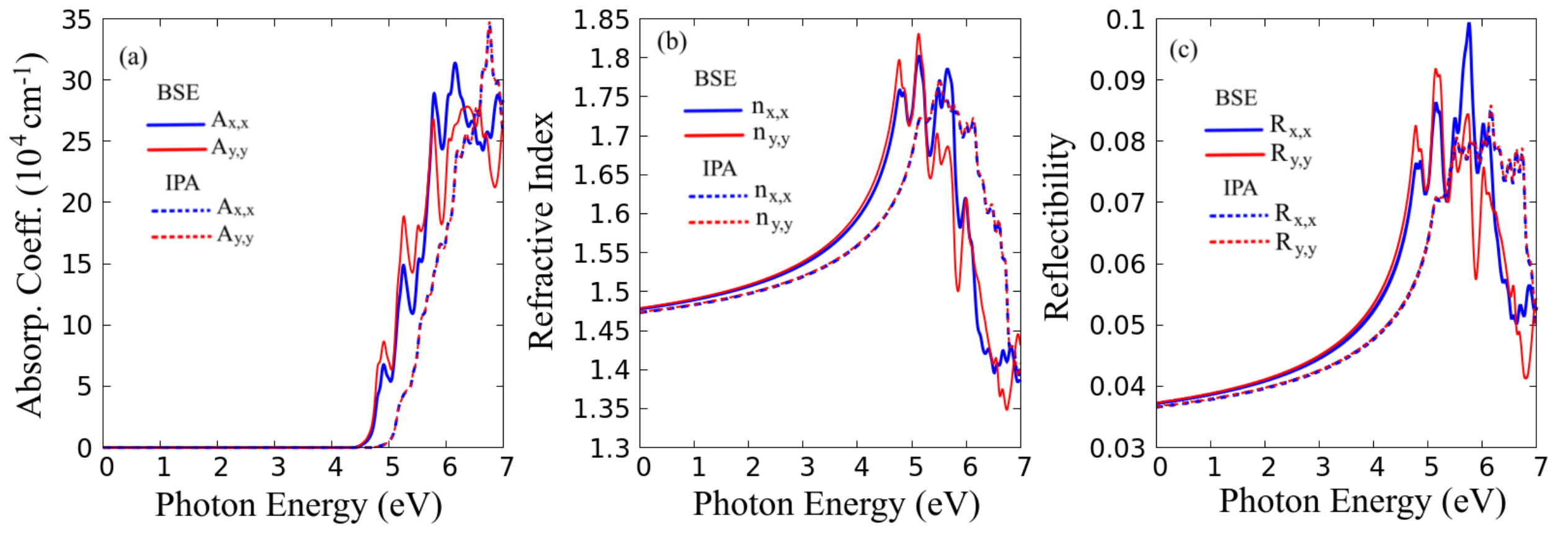}
    \caption{Optical properties of P4m2(1) phase of \ce{GeO2} monolayer, (a) refractive index and (b) reflectibility, at IPA (dashed curves) and BSE (solid curves) levels, considering a linear light polarization along $\hat{x}$ (blue curves) and $\hat{y}$ (red curves) directions.}
    \label{fig:opt_p4m2(1)}
\end{figure}

\begin{figure}[H]
    \centering
    \includegraphics[width=0.8\linewidth]{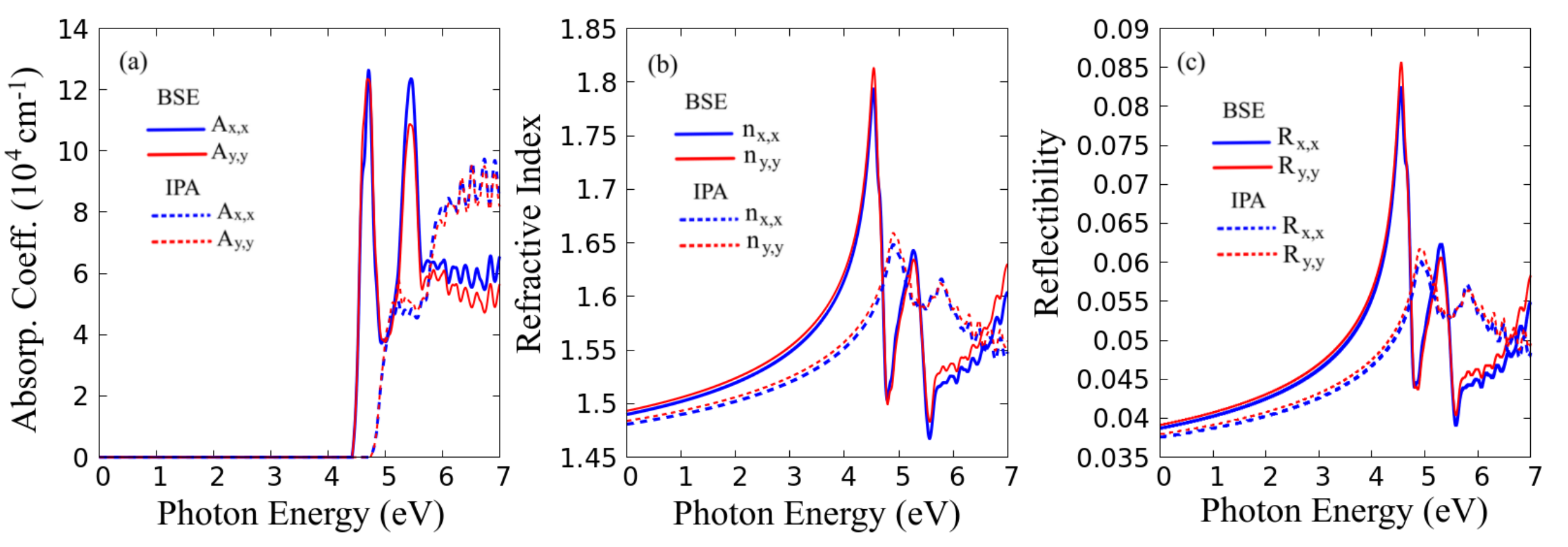}
    \caption{Optical properties of P4m2(2) phase of \ce{GeO2} monolayer, (a) refractive index and (b) reflectibility, at IPA (dashed curves) and BSE (solid curves) levels, considering a linear light polarization along $\hat{x}$ (blue curves) and $\hat{y}$ (red curves) directions.}
    \label{fig:opt_p4m2(2)}
\end{figure}

\begin{figure}[H]
    \centering
    \includegraphics[width=0.8\linewidth]{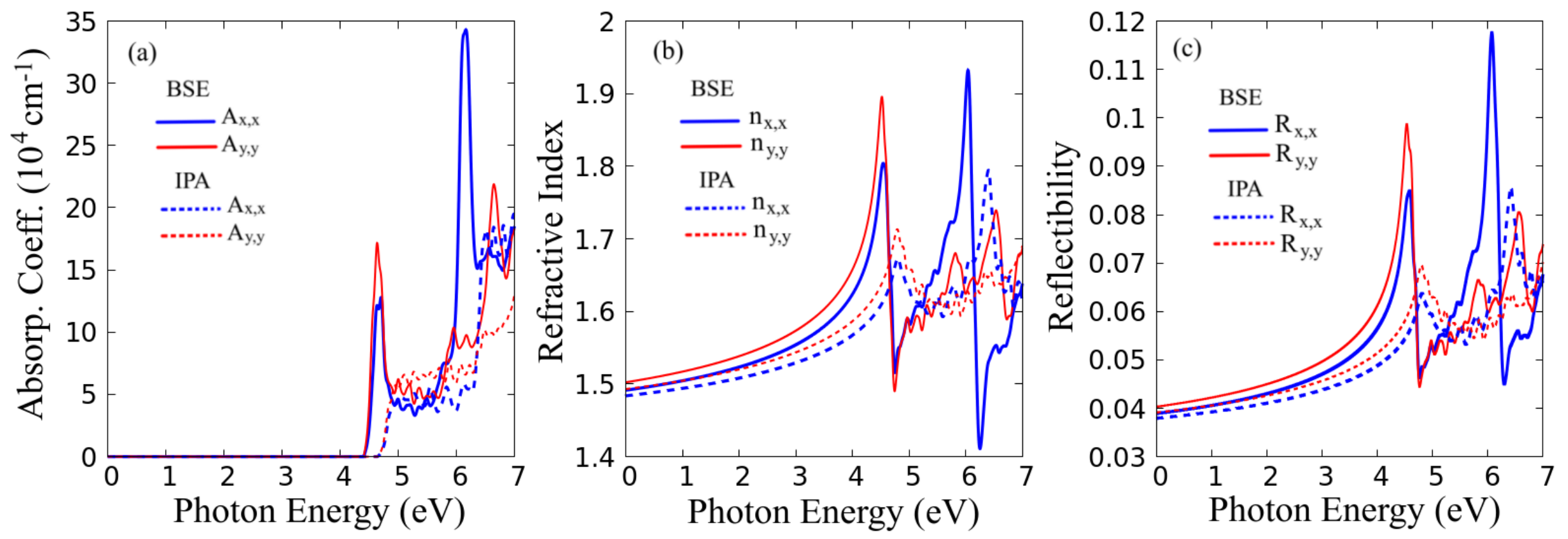}
    \caption{Optical properties of Pmmm(1) phase of \ce{GeO2} monolayer, (a) refractive index and (b) reflectibility, at IPA (dashed curves) and BSE (solid curves) levels, considering a linear light polarization along $\hat{x}$ (blue curves) and $\hat{y}$ (red curves) directions.}
    \label{fig:opt_pmmm(1)}
\end{figure}

\begin{figure}[H]
    \centering
    \includegraphics[width=0.8\linewidth]{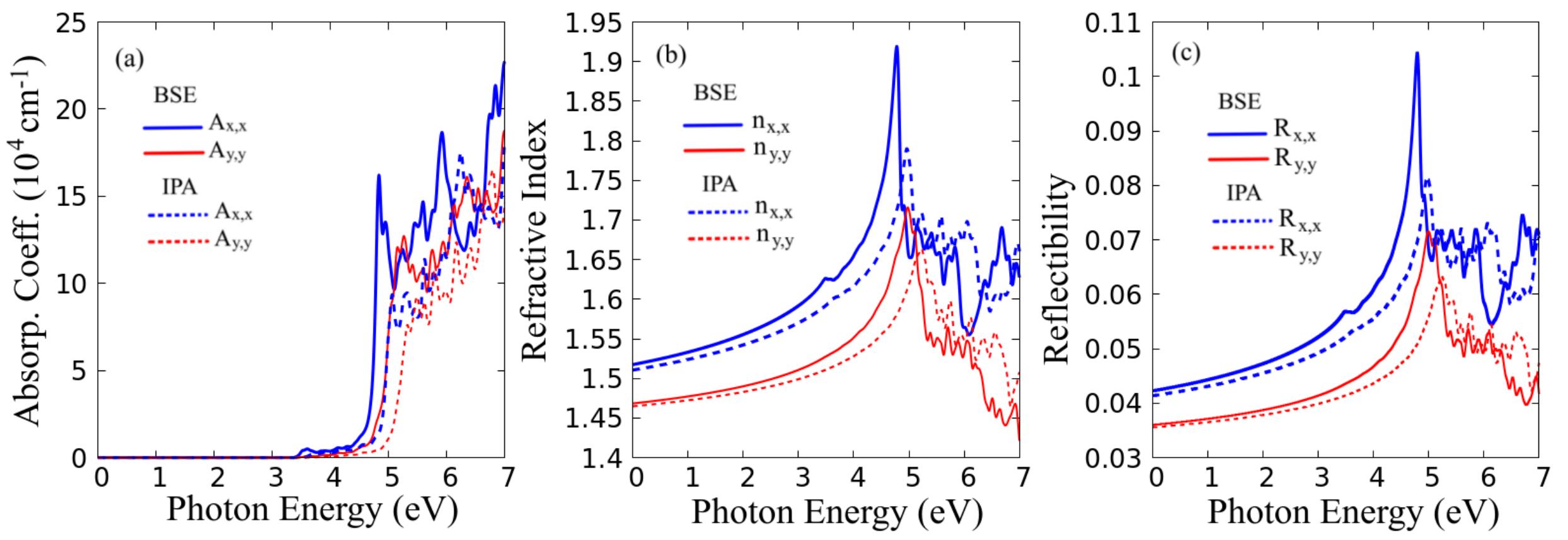}
    \caption{Optical properties of  \ce{GeO2} monolayer in the Pmmm(2) phase. (a) refractive index and (b) reflectibility, at IPA (dashed curves) and BSE (solid curves) levels, considering a linear light polarization along $\hat{x}$ (blue curves) and $\hat{y}$ (red curves) directions.}
    \label{fig:opt_pmmm(2)}
\end{figure}

\section{Band structure of P-3m1 including Ge-d orbitals}

\begin{figure}[H]
    \centering
    \begin{subfigure}{0.48\textwidth}
        \centering
        \includegraphics[width=\linewidth]{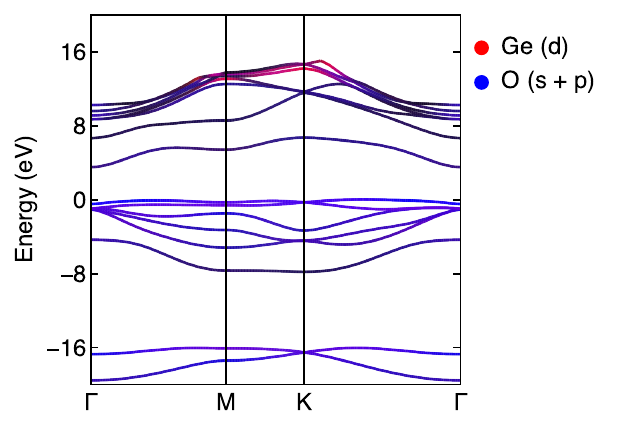}
        \caption{}
        \label{fig:ge_with_d}
    \end{subfigure}
    \hspace{0.1cm}
    \begin{subfigure}{0.49\textwidth}
        \centering
        \includegraphics[width=\linewidth]{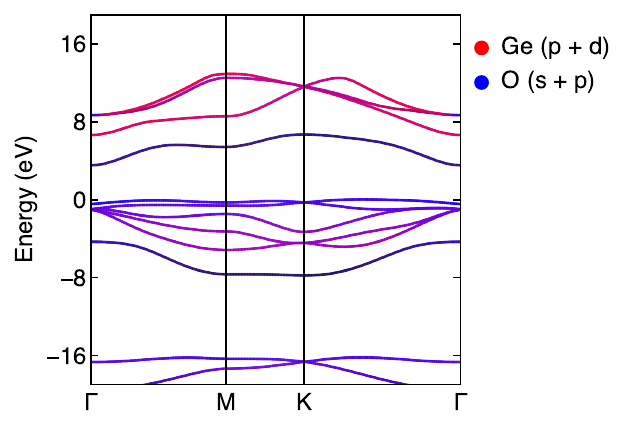}
        \caption{}
        \label{fig:ge_without_d}
    \end{subfigure}
    \caption{Projected band structure of the $P\text{-}3m1$ phase, calculated (a) considering the Ge $d$ orbitals as part of the valence set and (b) without them. The red markers indicate the contribution of Ge $d$ orbitals, while the blue markers show the contribution of O $s+p$ orbitals.}
    \label{fig:ge_bands}
\end{figure}

\section{Relaxed geometry of P6/mmm GeO$_2$ phase on Pt(111)(2x2) surface}

\begin{figure}[H]
\includegraphics[width=0.6\linewidth]{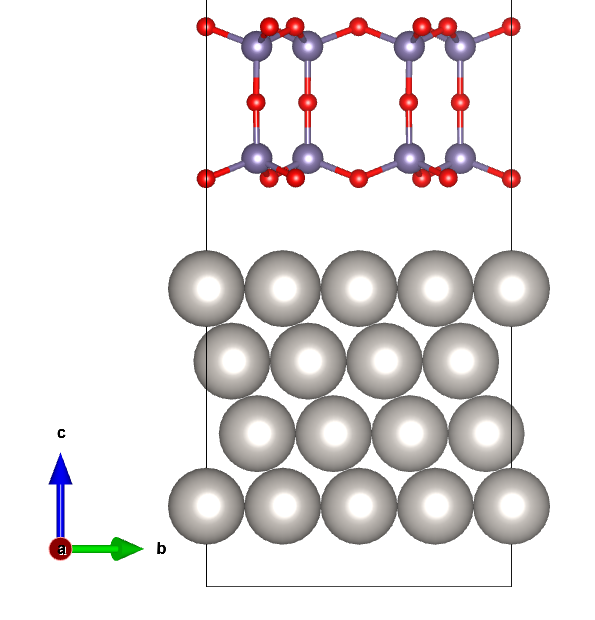}
\includegraphics[width=0.6\linewidth]{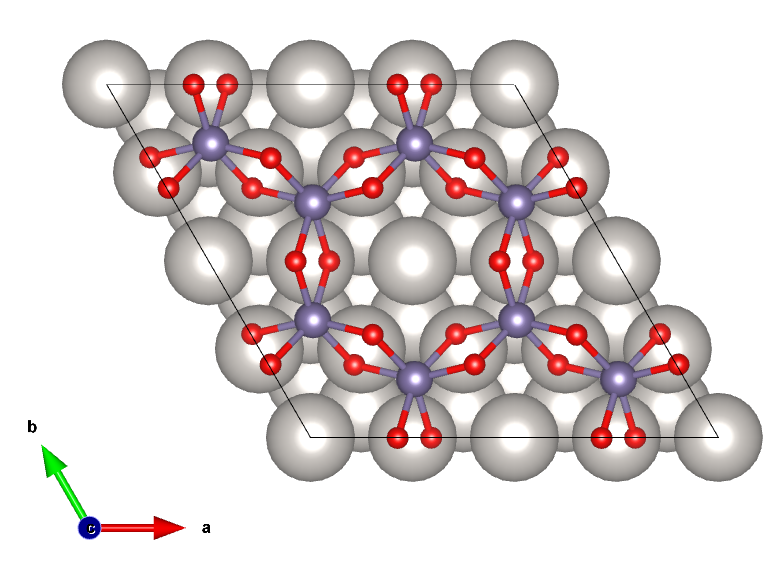}
\caption{Side and top view of relaxed geometry of P6/mmm/Pt(111)(2x2) surface. The surface lattice parameter of Pt(111)  is 11{\AA} in a slab model with 5 platinum layers, leading to a lattice parameter mismatch of 3.2\%.}
\label{fig:p6mmm_Pt111}
\end{figure}

\section{Absorption coefficient of P6/mmm Ge$_O$ phase with dielectric constant monolayered WSe$_2$ beneath}

\begin{figure}[H]
\includegraphics[width=0.6\linewidth]{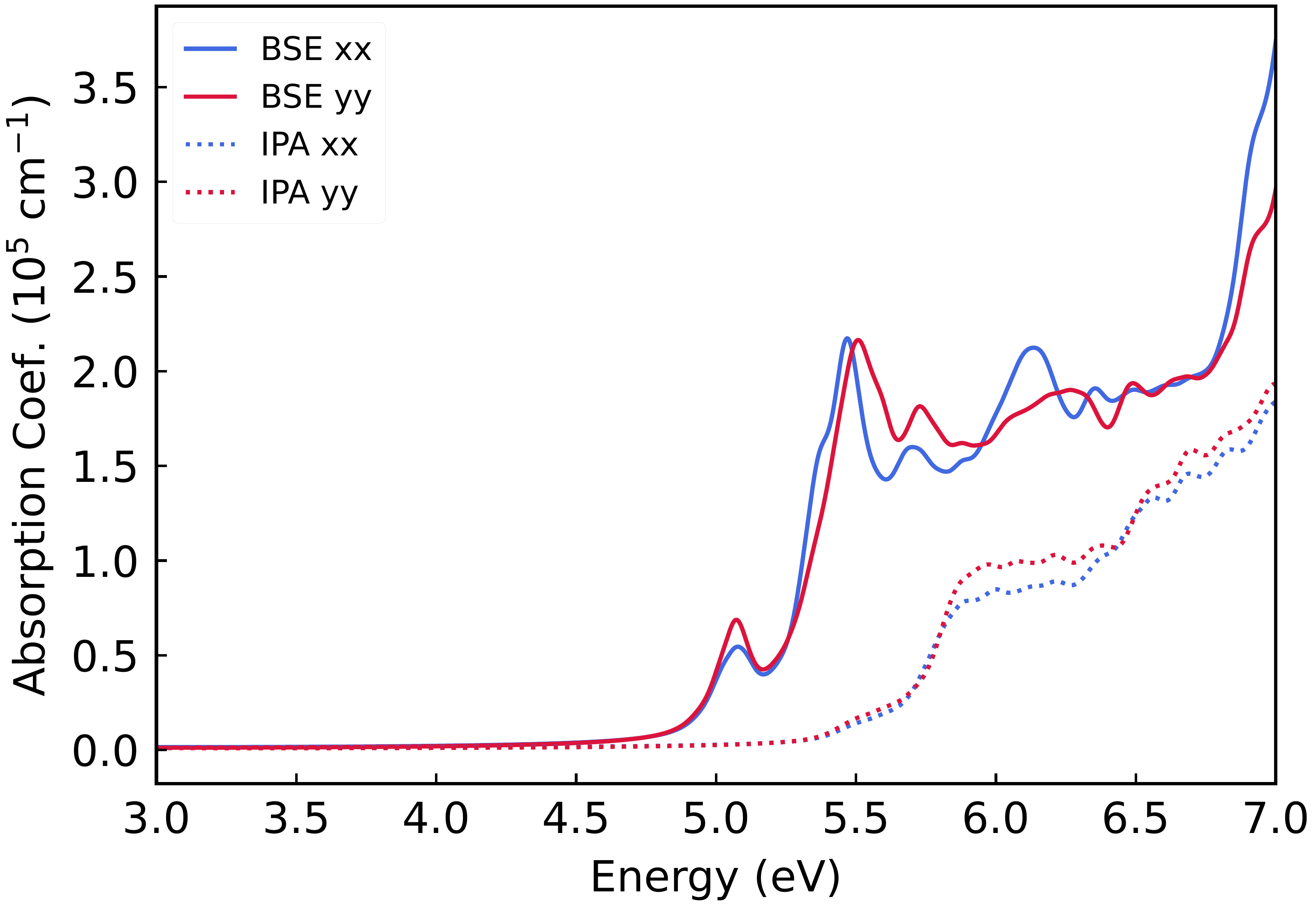}
\caption{Absorption coefficent of P6/mmm phase. The calculations have been performed using the dieletric constants of WSe$_2$ beneath the monolayer to mimic a substrate in contact with the P6/mmm GeO$_2$ phase.}
\label{fig:aimd}
\end{figure} 

\section{Equations for optical properties calculations}

The orthogonal TB Hamiltonian, $H(\mathbf{k})$, obtained within the Wannier90 framework can be described as:.

\begin{equation}
  \centering
H(\mathbf{k}) = H_{0}+ \sum_{i=1}^{N} e^{i \mathbf{k} \cdot \mathbf{R_{i}}} H_{\mathbf{R_{i}}}~,
\end{equation}

\noindent where $H_{0}$ corresponds to the Hamiltonian in unit cell, which contains the on-site energies and hopping parameters inside the cell. $H_{\mathbf{R_{i}}}$ corresponds to the hopping matrices, representing the interaction between unit cell and neighboring cells, while the matrix elements of $H_{0}$ and $H_{\mathbf{R_{i}}}$ are the output from the Wannier90 package. The electronic energy levels are obtained as: 

\begin{equation}
    H(\mathbf{k}) |n,\mathbf{k} \rangle = E_{n,\mathbf{k}} |n,\mathbf{k} \rangle,
\end{equation}

\noindent where $E_{n,\mathbf{k}}$ and $|n,\mathbf{k} \rangle$ are the eigenvalues and eigenvectors respectively, $n$ corresponds to the band index, $v$ for the valence (occupied) states and $c$ for the conduction (unoccupied) states at each $\textbf{k}$-point in the BZ.

To obtain the optical properties, we calculated the real and imaginary parts of the frequency dependent dielectric tensor, $\epsilon_{1,\alpha,\beta}(\omega)$ and $\epsilon_{2,\alpha,\beta}(\omega)$, respectively, through the following expressions:

\begin{align}
     \epsilon_{1,\alpha,\beta}(\omega) =  \delta_{\alpha,\beta} +  \frac{e^{2} S_{p}}{\epsilon_{0}\Omega N_{\mathbf{k}}} \sum_{\mathbf{k},c,v} F_{\alpha,\beta}^{c,v,\mathbf{k}} \frac{(E_{c,\mathbf{k}}-E_{v,\mathbf{k}})-\hbar \omega}{\left(\hbar \omega - (E_{c,\mathbf{k}}-E_{v,\mathbf{k}})\right)^{2}+\eta^{2}}~,
\\
   \epsilon_{2,\alpha,\beta}(\omega)=  \frac{ e^{2} S_{p}}{\epsilon_{0}\Omega N_{\mathbf{k}}} \sum_{\mathbf{k},c,v}  F_{\alpha,\beta}^{c,v,\mathbf{k}}   \frac{\eta}{\left(\hbar \omega - (E_{c,\mathbf{k}}-E_{v,\mathbf{k}})\right)^{2}+\eta^{2}}~,
\end{align}

\noindent where $\delta_{\alpha,\beta}$ is a kroenecker delta, $S_{p}$ is the spin factor (for spin and non-spin polarized cases). In the independent particle approximation (IPA), the oscillator force is given by:

\begin{equation}
    F_{\alpha,\beta}^{c,v,\mathbf{k}} =  \frac{\langle c,\mathbf{k} |P_{\alpha}|v,\mathbf{k} \rangle \langle v,\mathbf{k} |P_{\beta}|c,\mathbf{k} \rangle}{\left(E_{c,\mathbf{k}}-E_{v,\mathbf{k}}-i\eta_{1}\right) \left(E_{c,\mathbf{k}}-E_{v,\mathbf{k}}+i\eta_{1}\right)}~.
\end{equation}

\noindent $\Omega$ is the volume of unit cell, $\epsilon_{0}$ is the vacuum permittivity constant, $N_{\mathbf{k}}$ is the number of \textbf{k}-points employed for the BZ integration, $\omega$ is the incident photon frequency, $c(v)$ corresponds to the conduction (valence) states. $\eta$ is a parameter to smooth the dielectric function. $\alpha$ and $\beta$ correspond to the $x, y$ and z components in the dielectric tensor. $P_{\alpha}$ corresponds to the light-matter interaction operator given by $P_{\alpha} = \frac{\partial H(\mathbf{k})}{\partial k_{\alpha}}$. The absorption coefficient $A_{\alpha,\beta}(\omega)$ is then obtained as:

\begin{equation}
    A_{\alpha,\beta}(\omega)= \frac{\sqrt2\omega}{c}\left[\frac{\sqrt{\epsilon_{1,\alpha,\beta}^{2}(\omega)+\epsilon_{2,\alpha,\beta}^{2}(\omega)}-\epsilon_{1,\alpha,\beta}(\omega)}{2}\right]^{\frac{1}{2}},
\end{equation}

\noindent being $c$ the light speed. 

Finally we have calculated the absorption spectra within the Bethe--Salpeter equation (BSE) is a many-body equation that describes the interaction between an electron and a hole. The excitonic states are obtained through the solution of the two-particle problem via BSE (see main text). The exciton Hamiltonian, $H_{exc}$ is composed by the electron, $H_{e}$, and hole, $H_{h}$, single particle Hamiltonians plus the  Coulomb potential, $V_{eh}$, for the electron-hole pairs interaction $H_{exc} = H_{e} + H_{h}+V_{eh}~$.
The excitonic states with momentum center of mass $\mathbf{Q}$ can be expanded in terms of the product of electron and hole pairs wave functions as follows:

\begin{equation}
\Psi_{ex}^{n}(\mathbf{Q}) = \sum_{c,v,\mathbf{k}} A_{c,v,\mathbf{k},\mathbf{Q}}^{n} \ \left( |c,\mathbf{k}+\mathbf{Q} \rangle \otimes |v,\mathbf{k} \rangle \right)~, \label{eq:Exc_Basis}
\end{equation}

\noindent where the index $c$ and $v$ corresponds to the conduction and valence bands states, with momentum $\mathbf{k}+\mathbf{Q}$ and $\mathbf{k}$, respectively. The problem of excitons eigenvalues, can be transformed into BSE,  according to:

\begin{eqnarray}
    \left( E_{c,\mathbf{k}+\mathbf{Q}}-E_{v,\mathbf{k}}\right) A_{c,v,\mathbf{k},\mathbf{Q}}^{n}  +  \frac{1}{N_{k}}  \sum_{\mathbf{k'},v',c'} W_{(\mathbf{k},v,c),(\mathbf{k'},v',c'),\mathbf{Q}} \ A_{c',v',\mathbf{k'},\mathbf{Q}}^{n}  = E^{n}_{\mathbf{Q}} A_{c,v,\mathbf{k},\mathbf{Q}}^{n} \label{bse}~,
\end{eqnarray}

where $E^{n}_{\mathbf{Q}}$ are the energy of the excitonic state with momentum $\mathbf{Q}, A_{c,v,\mathbf{k},\mathbf{Q}}^{n}$ are the exciton wave function. $E_{c,\mathbf{k} + \mathbf{Q}} - E_{v,\mathbf{k}}$ are the single-particle energy difference between a conduction band state c with momentum $\mathbf{k}+\mathbf{Q}$ and a valence band state v with momentum $\mathbf{k}$, and $W_{(\mathbf{k},v,c),(\mathbf{k'},v',c'),\mathbf{Q}}$ are the many-body Coulomb interaction matrix element, which can be divided into two parts, direct interaction, $W^d$, and exchange interaction, $W^x$, respectively.

\begin{equation}
    W_{(\mathbf{k},v,c),(\mathbf{k'},v',c'),\mathbf{Q}}=W^d_{(\mathbf{k},v,c),(\mathbf{k'},v',c'),\mathbf{Q}}+W^x_{(\mathbf{k},v,c),(\mathbf{k'},v',c'),\mathbf{Q}}~.
\end{equation}

Since the Coulomb potential varies slightly inside unit cell in comparison with Bloch functions, we can approximate the orbital character of Coulomb term in the following way:

\begin{equation}
    W^{d}_{(\mathbf{k},v,c),(\mathbf{k'},v',c'),\mathbf{Q}}=V(\mathbf{k}-\mathbf{k'}) \ \langle c,\mathbf{k}+\mathbf{Q}|c',\mathbf{k'}+\mathbf{Q} \rangle \ \langle v',\mathbf{k'}|v,\mathbf{k} \rangle
\end{equation}
\noindent and $N_{k}$ is the number of $\textbf{k}$-points in the BZ. Within the BSE formalism, the real and imaginary parts of the frequency dependent dielectric tensor $\epsilon^{BSE}_{1,\alpha,\beta}(\omega)$ and $\epsilon^{BSE}_{2,\alpha,\beta}(\omega)$ are obtained as follows:

\begin{align}
\epsilon^{BSE}_{1,\alpha,\beta}(\omega) =  \delta_{\alpha,\beta} +  \frac{e^{2} S_{p}}{\epsilon_{0}\Omega N_{\mathbf{k}}} \sum_{n} F_{\alpha,\beta}^{n,BSE}
&\frac{E_{0}^{n}-\hbar \omega}{\left(\hbar \omega - E_{0}^{n}\right)^{2}+\eta^{2}}~,
\end{align}

\begin{align}
\epsilon^{BSE}_{2,\alpha,\beta}(\omega)=  \frac{e^{2} S_{p}}{\epsilon_{0}\Omega N_{\mathbf{k}}}  \sum_{n}  F_{\alpha,\beta}^{n,BSE}   \frac{\eta}{\left(\hbar \omega - E_{0}^{n}\right)^{2}+\eta^{2}}~,
\end{align}
\noindent where $E_{0}^{n}$ is the direct $(\mathbf{Q}=0)$ excitonic state energy, $F_{\alpha,\beta}^{n,BSE}$ is the excitonic oscillator force, calculated as:

\begin{align}
F_{\alpha,\beta}^{n,BSE} = \left( \sum_{c,v,\mathbf{k}} \frac{A^{n}_{c,v,\mathbf{k},0}\langle c,\mathbf{k} |P_{\alpha}|v,\mathbf{k} \rangle}{\left(E_{c,\mathbf{k}}-E_{v,\mathbf{k}}+i\eta_{1}\right)} \right) \left(\sum_{c',v',\mathbf{k'}} \frac{A^{n*}_{c',v',\mathbf{k'},0}\langle v',\mathbf{k'} |P_{\beta}|c',\mathbf{k'} \rangle}{\left(E_{c',\mathbf{k'}}-E_{v',\mathbf{k'}}-i\eta_{1}\right)} \right)\label{eq:bse-fosc} ~.
\end{align}

\end{document}